\documentclass{article}

\usepackage{graphicx} 
\usepackage[hidelinks]{hyperref} 
\usepackage{natbib} 
\usepackage{subcaption} 
\usepackage{enumitem} 
\usepackage{pifont} 
\usepackage{comment}
\usepackage[normalem]{ulem} 
 
\usepackage[skip=10pt plus1pt, indent=0pt]{parskip} 
\usepackage[margin=1in]{geometry} 

\usepackage{amssymb,amsmath,amsthm,amsfonts}
\usepackage{mathtools}
\usepackage{bbm}
\usepackage{color,xcolor}

\usepackage{multirow}

\newcommand{\one}{\mathbbm{1}}
\newcommand{\E}{\operatorname{\mathbbm{E}}}
\newcommand{\Q}{\operatorname{\mathbbm{P}}}
\newcommand{\R}{\mathbb{R}}

\newcommand{\Ff}{\mathcal{F}}

\newcommand{\Ii}{\mathcal{I}}

\newcommand{\Yy}{\mathcal{Y}}
\newcommand{\dd}{\: \mathrm{d}}

\DeclareMathOperator*{\argmin}{arg\,min}

\newtheorem{propo}{Proposition}

\theoremstyle{definition}
\newtheorem{defin}[propo]{Definition}
\newtheorem{remark}[propo]{Remark}

\title{Enforcing tail calibration when training probabilistic forecast models}
\date{}
\author{Jakob Benjamin Wessel\thanks{Department of Mathematics and Statistics, University of Exeter, Exeter, United Kingdom. \url{j.wessel@exeter.ac.uk.}} \and Maybritt Schillinger\thanks{Seminar for Statistics, ETH Zurich, Zurich, Switzerland. \url{maybritt.schillinger@stat.math.ethz.ch}} \and Frank Kwasniok\thanks{Department of Mathematics and Statistics, University of Exeter, Exeter, United Kingdom. \url{f.kwasniok@exeter.ac.uk.}} \and Sam Allen\thanks{Institute of Statistics, Karlsruhe Institute of Technology, Karlsruhe, Germany. \url{sam.allen@kit.edu}}}

\definecolor{darkteal}{rgb}{0,0.35,0.35}

\begin{document}

\maketitle

\begin{abstract}
    Probabilistic forecasts are typically obtained using state-of-the-art statistical and machine learning models, with model parameters estimated by optimizing a proper scoring rule over a set of training data. If the model class is not correctly specified, then the learned model will not necessarily issue forecasts that are calibrated. Calibrated forecasts allow users to appropriately balance risks in decision making, and it is particularly important that forecast models issue calibrated predictions for extreme events, since such outcomes often generate large socio-economic impacts. In this work, we study how the loss function used to train probabilistic forecast models can be adapted to improve the reliability of forecasts made for extreme events. We investigate loss functions based on weighted scoring rules, and additionally propose regularizing loss functions using a measure of tail miscalibration. We apply these approaches to a hierarchy of increasingly flexible forecast models for UK wind speeds, including simple parametric models, distributional regression networks, and conditional generative models. We demonstrate that state-of-the-art models do not issue calibrated forecasts for extreme wind speeds, and that the calibration of forecasts for extreme events can be improved by suitable adaptations to the loss function during model training. This introduces a trade-off between calibrated forecasts for extreme events and calibrated forecasts for more common outcomes.
\end{abstract}

\section{Introduction}\label{sec:intro}


Probabilistic forecasts take the form of probability distributions over the set of possible outcomes. Such forecasts comprehensively describe the uncertainty in the unknown outcome, making them essential for effective risk assessment and decision making. They have therefore become commonplace in a variety of application domains, including, for example, medicine, economics, finance, politics, and climate science.


Probabilistic forecasts are typically obtained using state-of-the-art statistical and machine learning models, which learn the conditional distribution of the outcome variable given some covariates. Parameters of the models can be estimated by optimizing a \emph{proper scoring rule} over a set of training data. Proper scoring rules quantify forecast accuracy \citep[see e.g.][]{GneitingRaftery2007}, and this framework therefore finds the model parameters that result in the most accurate forecasts, with accuracy measured in terms of the chosen scoring rule. If the model class is well-specified (i.e. the family of parametrized candidate models contains the true conditional distribution of the outcome), then this \emph{optimum score estimation} should return the true parameters underlying the data generating process, provided sufficient data is available \citep{GneitingRaftery2007,DawidEtAl2016}. This holds for any choice of proper scoring rule.


However, if the model class is not correctly specified, as is generally the case in practice, then the resulting forecasts will exhibit biases compared to the true distribution underlying the data. These biases will generally change for different scoring rules. For example, it is well-known that model estimation with the continuous ranked probability score \citep[CRPS;][]{MathesonWinkler1976}, a popular proper scoring rule, tends to yield forecast distributions that are less dispersed than those obtained from maximum likelihood estimation \citep[see e.g.][]{GneitingEtAl2005,gebetsberger_estimation_2018,BuchweitzEtAl2025}. 

In this mis-specified case, there is therefore no guarantee that the estimated model yields forecasts that are \emph{calibrated} (or \textit{reliable}). Probabilistic forecasts are calibrated if they align statistically with the corresponding outcomes, which is typically assessed by checking whether forecast probability integral transform (PIT) values resemble a sample from a standard uniform distribution \citep{Dawid1984,DieboldEtAl1998,GneitingEtAl2007}. \cite{GneitingEtAl2007} remarked that the general goal of probabilistic forecasting is to issue forecasts that are as informative as possible, subject to being calibrated. In this sense, calibration is a necessary requirement that forecasts must satisfy if they are to be useful for decision making and risk assessment. 


This is particularly true when forecasting extreme outcomes. Extreme events typically lead to the largest impacts on forecast users, and calibrated forecasts for such outcomes are therefore particularly valuable. However, \cite{AllenEtAl2024} demonstrate that a forecast can satisfy standard notions of calibration without issuing reliable forecasts for extreme events. They therefore introduce a notion of \emph{tail calibration} that assesses whether forecasts are calibrated when predicting extremes. This facilitates a more comprehensive understanding of how forecasts behave when interest is on extreme outcomes, helping practitioners to design forecast models that are simultaneously calibrated when predicting extreme and non-extreme events. 


One approach to obtain more accurate forecasts for extreme events is to train probabilistic forecast models by optimizing a weighted scoring rule \citep{WesselEtAl2024}. Weighted scoring rules allow users to target particular outcomes when quantifying forecast accuracy, and are therefore commonly employed to compare forecasts for extreme outcomes \citep{LerchEtAl2017}. However, there is again no guarantee that models trained using a weighted scoring rule will issue forecasts that are tail calibrated. As \cite{Wilks2018} remark, ``calibration is not guaranteed because [proper score] minimization may be achieved as a compromise between calibration and resolution (which is related to sharpness)''. The same holds when interest is on weighted scoring rules and tail calibration.


As an alternative, one could regularize the scoring rule so that it penalizes (tail) miscalibration directly. \cite{Wilks2018}, for example, propose measuring to what extent the distribution of PIT values deviates from a standard uniform distribution, and including this measure of miscalibration as an additional term in the loss function, thereby encouraging forecasters to issue calibrated forecasts. The resulting forecasts are typically better calibrated than those obtained directly from optimum score estimation, though this may come at the expense of forecast accuracy or sharpness. If interest is on extreme events, a similar framework could be adopted, in which one measures the tail miscalibration of the forecasts, and then adds this to the scoring rule during model training. 


In this work, we investigate how regularizing the loss function using either weighted scoring rules or measures of miscalibration affects the performance of the resulting forecasts when interest is on extreme outcomes. These different regularization terms are implemented in a case study on the statistical post-processing of weather forecasts. We consider a hierarchy of increasingly more flexible probabilistic wind speed models, including simple baseline parametric methods \citep{GneitingEtAl2005}, distributional regression networks \citep{rasp_neural_2018}, and conditional generative models \citep{chen_generative_2024}. We first evaluate the performance of these methods with respect to extreme outcomes, and demonstrate that all three approaches fail to issue forecasts that are tail calibrated; increasing the flexibility of the model class does not solve this. We then demonstrate how changing the loss function affects the resulting forecast performance. Penalizing tail miscalibration can improve the reliability of the resulting forecasts for extreme events, though this comes at the expense of overall forecast calibration and performance.


The remainder of the paper is structured as follows. The following section introduces probabilistic forecasts, proper scoring rules, and the theory of optimum score estimation. Section \ref{sec:calibration} defines the notions of calibration and tail calibration that we want our forecasts to satisfy, and Section \ref{sec:enforce} then proposes different regularization terms that could be employed during model training to weakly enforce (tail) calibration of the resulting probabilistic forecasts. A simple demonstration of the approach on simulated data is presented in Section \ref{sec:simstudy}, before an application to statistical weather forecasting models is discussed in Section \ref{sec:applications}. Section \ref{sec:conclusion} concludes.

\section{Training probabilistic forecast models}\label{sec:weightedscores}

\subsection{Proper scoring rules}

Let $Y \in \Yy \subseteq \R$ denote the outcome variable that we wish to forecast. Here, a probabilistic forecast for $Y$ is a probability distribution on the outcome space $\Yy$. We use $\Ff$ to denote a class of such distributions, and represent a probabilistic forecast $F \in \Ff$ via its cumulative distribution function. 

A scoring rule is a function $S : \Ff \times \Yy \to \R \cup \{-\infty, \infty\}$ that takes a probabilistic forecast $F \in \Ff$ and an observation $y \in \Yy$ as inputs, and outputs a real-valued (possibly infinite) score $S(F, y)$ that quantifies the forecast's accuracy. A lower score corresponds to a more accurate forecast, and a scoring rule is called \emph{proper} if
\[
\E_{Y \sim G}[S(G, Y)] \leq \E_{Y \sim G}[S(F, Y)] \quad \text{for all} \quad F, G \in \Ff.
\]
That is, if the outcome follows the true distribution $G \in \Ff$, then the score is minimized in expectation by issuing $G$ as the forecast. The scoring rule is \emph{strictly proper} if this holds with equality if and only if $F = G$.

Proper scoring rules condense forecast performance into a single value, providing a means to rank and compare competing forecasts. While all proper scoring rules are minimized in expectation by the true distribution of the outcomes, the ranking of imperfect forecasts will generally depend on the chosen scoring rule. Popular examples of proper scoring rules for real-valued outcomes include the logarithmic (log) score, 
\begin{equation}\label{eq:logs}
    \text{LS}(F, y) = -\log f(y),
\end{equation}
where $f$ is the probability density function associated with $F$, assuming it exists \citep{Good1952}; and the continuous ranked probability score (CRPS),
\begin{equation}\label{eq:crps}
    \text{CRPS}(F, y) = \int_{-\infty}^{\infty} \left[F(z) - \one\{y \leq z\}\right]^{2} \dd z,
\end{equation}
where $\one\{ \cdotp \}$ denotes the indicator function \citep{MathesonWinkler1976}. We refer to \cite{GneitingRaftery2007} and \citet{WaghmareZiegel2025} for comprehensive overviews on proper scoring rules.

\subsection{Optimum score estimation}

Probabilistic forecasts for $Y$ are generally obtained using a statistical or machine learning model that estimates the conditional distribution of $Y$ given some covariates. These probabilistic forecast models depend on parameters $\theta \in \Theta \subseteq \R^{d}$ that link the covariates to the outcome, and we denote the forecast distribution corresponding to parameter vector $\theta$ and covariate vector $x$ by $F_{\theta, x}$. We say that a model is well-specified if there exists a parameter vector $\theta^{*} \in \Theta$ for which, for any covariates $x$, the resulting forecast distribution $F_{\theta^{*}, x}$ is equal to the conditional distribution of $Y$ given the covariates.

In practice, we typically have access to a (training) set of covariates and outcomes $\{(x_{1}, y_{1}), \dots, (x_{n}, y_{n})\}$ from which to learn the optimal model parameters. Since forecast accuracy is assessed using proper scoring rules, it makes sense to find the parameter vector $\theta$ that optimizes the average score over the training data. This is generally referred to as optimum or minimum score estimation \citep{GneitingRaftery2007}. 
\\
\begin{defin}
    The optimum score estimator corresponding to a proper scoring rule $S$ is
    \begin{equation}\label{eq:ose}
       \hat{\theta} = \argmin_{\theta \in \Theta} \frac{1}{n} \sum_{i=1}^{n} S(F_{\theta,x_i}, y_{i}).
    \end{equation}
\end{defin}

Optimizing the log score is equivalent to maximum likelihood estimation, while many studies have also considered using the CRPS to estimate model parameters \citep[see e.g.][]{GneitingEtAl2005}. The inferential properties of optimum score estimators have been studied in depth by \cite{DawidEtAl2016}. 

If $S$ is a strictly proper scoring rule, then the optimum score estimator is consistent for $\theta^{*}$ \citep{GneitingRaftery2007}. Hence, if the forecast model is well-specified, then optimum score estimation with any strictly proper scoring rule will asymptotically recover the parameter vector $\theta^{*}$ corresponding to the true conditional distribution of $Y$ given the covariates. In this case, the choice of strictly proper scoring rule to employ in \eqref{eq:ose} is largely insignificant. However, in practice, forecasting models are generally not correctly specified -- it is common to assume a particular parametric distribution for the response variable, for example -- and we only have access to a finite amount of data. In most practical applications, minimizing the average score will therefore not recover the true distribution underlying the outcomes. Hence, employing different scoring rules within \eqref{eq:ose} will generally result in different parameter estimates.

\section{Tail calibration}\label{sec:calibration}

Having estimated the optimal parameters of a probabilistic forecasting model, the fit of the model can be assessed by checking whether or not the resulting forecast distributions are calibrated. Loosely speaking, forecasts are calibrated if they can be taken at face value, which is necessary for effective decision making. 

To define calibration formally, let $F$ now denote a random cumulative distribution function, acting as a probabilistic forecaster. While $F$ often depends on parameters and (random) covariates, as discussed above, we drop this dependence from the notation in this section for ease of presentation. The forecast-observation pairs $\{(F_{1}, y_{1}), \dots, (F_{n}, y_{n})\}$ that we observe in practice can be interpreted as realisations of the random pair $(F, Y)$, and we use the probability measure $\Q$ to describe the joint distribution of $F$ and $Y$. This corresponds to the prediction space setting introduced by \cite{GneitingRanjan2013}, to which readers are referred for further details. Since we are interested in extremes, we assume that $F$ is almost surely continuous, though simple adaptations to the definitions below are available when the forecast distributions are not continuous \citep[e.g.][Definition 2.5]{GneitingRanjan2013}. We study the following notion of forecast calibration \citep[see][]{Dawid1984,DieboldEtAl1998,GneitingEtAl2007}.
\\
\begin{defin}
    A forecaster is \emph{probabilistically calibrated} if the forecast probability integral transform (PIT) $F(Y)$ follows a standard uniform distribution, that is,
    \begin{equation}\label{eq:prob_cali}
        \Q(F(Y) \le u) = u \quad \text{for all} \quad u \in [0, 1].
    \end{equation}
\end{defin}

Probabilistic calibration can be assessed in practice by checking whether the empirical PIT values $\{F_1(y_1), \dots, F_n(y_n)\}$ resemble a sample from a standard uniform distribution, which is typically performed by plotting a histogram or pp-plot of the PIT values. The shape of the diagnostic plot can provide useful information regarding what biases occur in the forecast. 

As \cite{GneitingRanjan2013} remark, ``checks for the uniformity of the probability integral transform have formed a cornerstone of density forecast evaluation.'' However, probabilistic calibration is a fairly weak notion of calibration, since it concerns unconditional facets of the forecasts and observations \citep[e.g.][]{GneitingResin2023}. \cite{AllenEtAl2024} demonstrate that a forecaster can be probabilistically calibrated without necessarily issuing reliable forecasts for extreme events, and therefore introduce notions of forecast \emph{tail calibration}, which assess the calibration of forecasts when interest is only in the tails of the predictive distribution. 

Given a threshold $t \in \R$, define the forecast excess distribution as
\begin{equation}
F_t(x) = \frac{F(x) - F(t)}{1 - F(t)}, \quad x \geq t, 
\end{equation}
with $F_t(x) = 0$ for $x < t$, and $F_t(x) = 1$ for all $x \in \R$ if $F(t) = 1$. If $Y \sim G$, then the excess distribution $G_t$ describes the distribution function of $Y$ conditioned on $Y$ exceeding the threshold $t$. The term ``excess distribution'' is also often used to describe the conditional distribution of $Y - t$ given $Y > t$, but we employ the definition above here since it simplifies the notation below. 
\\
\begin{defin}\label{def:ptc}
    A forecaster $F$ is \emph{probabilistically tail calibrated} at a threshold $t \in \R$ if
    \begin{equation}\label{eq:ptc2}
        \Q(Y > t) = \E[1 - F(t)],
    \end{equation}
    and 
    \begin{equation}\label{eq:ptc3}
    \Q(F_t(Y) \le u \mid Y > t) = u \quad \text{for all $u \in [0, 1]$.}
    \end{equation}
\end{defin}
\bigskip
\begin{remark}
    This differs slightly from the definition of tail calibration in \cite{AllenEtAl2024}, who consider the limit as $t$ tends to the upper end point of the distribution of $Y$. Since both approaches require assessing tail calibration at a finite threshold (or thresholds) in practice, this difference is largely irrelevant for this study.
\end{remark}

The first condition in Definition \ref{def:ptc} focuses on the forecast threshold exceedance probability, thereby assessing forecasts for the occurrence of an extreme event, while the second condition concerns the calibration of the conditional forecast distribution given that the threshold $t$ has been exceeded, thereby assessing forecasts for the severity of extreme events given their occurrence. The latter condition is equivalent to assessing whether the forecast excess distribution is probabilistically calibrated when predicting outcomes that exceed the threshold of interest, which has been suggested by \cite{AllenEtAl2023b} and \cite{MitchellWeale2023}. This can be assessed empirically via conditional PIT (CPIT) values, 
\[z_{i,t} = F_{i, t}(y_i)\text{, for } i \in \Ii_{t},\]
where $\Ii_{t} = \{ i \in \{ 1, \dots, n \} : y_{i} > t \}$ is the set of indices for which the outcome exceeds the threshold, and $F_{i,t}$ is the excess distribution of forecast $F_i$ at threshold $t$. If a histogram of the empirical CPIT values appears flat, or a pp-plot lies close to the diagonal, there is evidence to suggest that \eqref{eq:ptc3} is satisfied.

The CPIT values only assess the shape of the forecast distribution above the threshold, and ignore the probability with which threshold exceedances are forecast to occur; this is captured separately by \eqref{eq:ptc2}. The two conditions of Definition \ref{def:ptc} can alternatively be combined into one criterion that incorporates both the shape of the excess distribution and the occurrence of threshold exceedances.
\\
\begin{propo}
    Let $O_t = \Q(Y > t) / \E[1 - F(t)]$, and let $H_{Z_t}$ denote the distribution function of $F_t(Y)$ given that $Y > t$. If $H_{Z_t}$ is strictly increasing, with inverse $H_{Z_t}^{-1}$, then the two conditions in Definition \ref{def:ptc} hold if and only if
    \begin{equation}\label{eq:Q}
        O_t H_{Z_t}^{-1}(u) = u \quad \text{for all $u \in [0, 1]$.}
    \end{equation}
\end{propo}

\begin{proof}
    Equation \eqref{eq:ptc2} is recovered from \eqref{eq:Q} when $u = 1$, and \eqref{eq:ptc3} follows by rearranging \eqref{eq:Q}. The opposite direction, that \eqref{eq:ptc2} and \eqref{eq:ptc3} imply \eqref{eq:Q}, is trivial.
\end{proof}

\cite{AllenEtAl2024} suggest an alternative combination of \eqref{eq:ptc2} and \eqref{eq:ptc3} (namely, $O_tH_{Z_t}(u) = u$ for all $u \in [0, 1]$), but we argue in Appendix \ref{app:estimators} that this can be less interpretable than the expression in \eqref{eq:Q}, and is less suited to the methodology proposed in the following section.

To assess whether \eqref{eq:Q} holds in practice, we can estimate $O_tH_{Z_t}^{-1}$ using 
\begin{equation}\label{eq:decomposition}
    \hat{Q}_t(u) = \underbrace{\frac{\lvert \Ii_t \rvert}{\sum_{i=1}^n (1 - F_i(t))}}_{\hat{O}_t} \cdot \underbrace{\vphantom{\frac{\lvert \Ii_t \rvert}{\sum_{i=1}^n [1 - F_i(t)]}} \hat{q}_{z_t}(u)}_{\hat{H}_{Z_t}^{-1}(u)} \qquad \text{for $u \in [0, 1]$},
\end{equation}
where $\hat{q}_{z_t}(u)$ is the sample $u$-quantile derived from the conditional PIT values $z_{i,t}$, $i \in \Ii_t$. If $\hat{Q}_t(u)$ is close to $u$ for all $u \in [0, 1]$, then there is evidence to suggest that the forecasts are probabilistically tail calibrated. In this case, a plot of $\hat{Q}_t(u)$ against $u$ should therefore lie close to the diagonal. In the limit as $t \to -\infty$, probabilistic tail calibration recovers probabilistic calibration: condition \eqref{eq:ptc2} then becomes trivial, condition \eqref{eq:ptc3} coincides with \eqref{eq:prob_cali}, and the estimate $\hat{Q}_t$ is equal to the empirical quantile function of the standard PIT values. Plotting $\hat{Q}_t(u)$ against $u \in [0, 1]$ is then equivalent to a qq-plot of the PIT values.

Probabilistic calibration does not imply probabilistic tail calibration (or vice versa), meaning if the standard checks for probabilistic calibration are satisfied, this does not necessarily ensure that the forecasts are calibrated in the tail \citep{AllenEtAl2024}. One consequence of this is that probabilistic forecast models may result in forecasts that are calibrated but not tail calibrated. In the following, we discuss how loss functions can be adapted so that tail calibration is enforced during model training.

\section{Enforcing tail calibration}\label{sec:enforce}

While proper scoring rules implicitly incorporate forecast calibration when quantifying forecast performance, training incorrectly specified forecast models by optimizing a proper scoring rule will not necessarily yield calibrated predictions. Proper scores can be decomposed into terms of discriminative ability and miscalibration \citep{Murphy1973,Brocker2009}, and the learned forecast distributions will not necessarily result in a miscalibration term that is equal to zero since both calibration and sharpness are optimized; the scoring rule might prefer forecast distributions that are more informative, even if they are miscalibrated.

Moreover, even if the resulting forecasts appear probabilistically calibrated, there is no guarantee that they also yield calibrated forecasts for extreme events. In this section, we therefore discuss possible ways to regularize scoring rule-based loss functions during model training so that tail calibration of the resulting forecasts is encouraged. We use the terms regularization and penalization interchangeably. Table \ref{tab:loss_fct} provides an overview of the loss functions that we will study.

\renewcommand{\arraystretch}{1.2}
\begin{table}
\caption{Loss functions studied in this work. Here, $\bar{S} = \tfrac{1}{n}\sum_{i=1}^{n}S(F_i, y_i)$ denotes the average score corresponding to some strictly proper scoring rule $S$, such as the log score or CRPS, and $\bar{S}_w = \tfrac{1}{n} \sum_{i=1}^{n}S(F_i, y_i; w)$ denotes the average score corresponding to some weighted scoring rule, such as the censored likelihood score or threshold-weighted CRPS.}
\begin{tabular}{|l|l|c|l|}
\hline
Name     & Loss function                          & Equation & Interpretation            \\ \hline
Baseline & $\bar{S}$             & \eqref{eq:ose} with \eqref{eq:logs} or \eqref{eq:crps} & Optimum score estimation with no penalization.                          \\ 
Weighted score & $\bar{S} + \gamma \bar{S}_{w}$ & \eqref{eq:ose} with \eqref{eq:cls} or \eqref{eq:twcrps}    & Optimum weighted score estimation.     \\
MCB      & $\bar{S} + \gamma \text{MCB}$      & \eqref{eq:regu}       & Penalization of probabilistic miscalibration. \\
TMCB     & $\bar{S} + \gamma \text{TMCB}$     & \eqref{eq:tregu}       & Penalization of probabilistic tail miscalibration.          \\ \hline
\end{tabular}
\label{tab:loss_fct}
\end{table}

\subsection{Weighted scoring rules}

In the case of an incorrectly specified forecast model, \cite{GneitingRaftery2007} argue that ``the appeal of optimum score estimation lies in the potential adaption of the scoring rule to the problem at hand.'' That is, proper scoring rules can be used to target particular aspects of forecast distributions when quantifying forecast accuracy, and by using these scoring rules as loss functions, the resulting model should generate forecasts that accurately predict these aspects of interest. 

To target particular outcomes whilst quantifying forecast accuracy, scoring rules can be adapted to incorporate a weight function $w: \Yy \to [0, 1]$, which can be selected such that higher weight is assigned to outcomes of more interest. This recognizes that accurate forecasts for certain outcomes, such as extreme events, are often more valuable than forecasts of other outcomes. 

\cite{DiksEtAl2011} introduce the censored likelihood score, a weighted version of the log score, as
\begin{equation}\label{eq:cls}
    \text{cLS}(F, y; w) = -w(y) \log f(y) + [1 - w(y)] \log \left[ 1 - \int_{-\infty}^{\infty} w(z) f(z) \dd z \right].
\end{equation}
Similarly, \cite{GneitingRanjan2011} introduce the threshold-weighted CRPS,
\begin{equation}\label{eq:twcrps}
    \text{twCRPS}(F, y; w) = \int_{-\infty}^{\infty} \left[F(z) - \one\{y \leq z\}\right]^{2} w(z) \dd z.
\end{equation}
The log score and CRPS are recovered when $w(z) = 1$ for all $z \in \Yy$. 

To put emphasis on values above a threshold $t \in \R$, it is common to employ the weight function $w(z) = \one\{z \geq t\}$. In this case, the censored likelihood score and twCRPS both censor the forecast distribution and the outcome at the threshold, before applying the relevant unweighted scoring rule. This framework can readily be applied to other scoring rules, also in a multivariate context \citep{AllenEtAl2023,dePunderEtAl2023}. \cite{HolzmannKlar2017} propose an alternative framework to construct weighted scoring rules that separately assess the conditional forecast distribution given that the threshold has been exceeded, and the forecast probability of a threshold exceedance; this encompasses the censored likelihood score but not the twCRPS.

Since weighted scoring rules allow events of interest to be targeted during forecast evaluation, they can similarly be used to target events of interest during model training. The resulting forecasts should issue more accurate predictions for these events. \cite{WesselEtAl2024} propose training statistical post-processing methods for weather forecasts by using weighted scoring rules as loss functions. They find that doing so tends to decrease overall forecast performance, as assessed using an unweighted scoring rule, but improves forecast performance in the region of interest. This illustrates the trade-off between generating accurate forecasts for outcomes that more commonly occur, and accurate forecasts for high-impact events.

If the weight function used in \eqref{eq:cls} and \eqref{eq:twcrps} is of the form $w(z) = \one\{z \geq t\}$, then the censored likelihood score and twCRPS are generally proper but not strictly proper; the scoring rules are insensitive to how the forecast distributions behave below the threshold. This can become problematic during model training since, depending on the class of models being trained, it can lead to non-unique solutions to the optimization problem. Instead, we consider a linear combination of weighted and unweighted scoring rules. For example, for the CRPS, we employ
\begin{equation}
\text{CRPS}(F, y) + \gamma \text{twCRPS}(F, y; w),
\end{equation}
where $\gamma \geq 0$ represents the amount of emphasis placed on the weighted scoring rule, and the weight function in the twCRPS is $w(z) = \one\{z \geq t\}$. This combined score itself corresponds to the twCRPS with weight function $w(z) = 1 + \gamma \one\{z \geq t\}$, which has been proposed for evaluation purposes by \cite{LerchEtAl2017} and \cite{vannitsem_2018_chapter_6}. Whenever the CRPS is strictly proper, this linear combination will also be strictly proper. An analogous combination can be constructed using the log score and censored likelihood score, or any other scoring rule and a weighted counterpart. Note that the scale of the weighted and unweighted scores will generally differ, and hence $\gamma$ has no direct interpretation; it must therefore be chosen using some exploratory analysis. In Sections \ref{sec:simstudy} and \ref{sec:applications}, we explore how $\gamma$ influences the forecast performance.

\subsection{Calibration and tail calibration regularization}\label{sec:penalty}

In order to direct optimum score estimators towards a calibrated solution, \cite{Wilks2018} suggest introducing a measure of miscalibration (MCB), and using this as a regularization term within the loss function. That is, to train the forecasting model by minimizing
\begin{equation}\label{eq:regu}
    \frac{1}{n} \sum_{i=1}^{n} S(F_{i}, y_{i}) + \gamma \text{MCB},
\end{equation}
where $\gamma \geq 0$ represents the amount of emphasis placed on the regularization term, with $\gamma = 0$ recovering the un-regularized optimum score estimator. The miscalibration term MCB is implicitly a function of the training data $\{(F_{1}, y_{1}), \dots, (F_{n}, y_{n})\}$. By adding the regularization term, the resulting forecasts should exhibit better calibration, though possibly with a worse average score. This trade-off is amplified as the parameter $\gamma$ is increased.

An obvious question is what miscalibration term to use. Since the loss function will be evaluated multiple times whilst being optimized, MCB should be easy to calculate. \cite{Wilks2018} consider a measure of miscalibration that is specific to discrete forecast distributions, but in principle, any divergence between the empirical distribution of the $n$ PIT values $\{z_1 = F_{1}(y_{1}),..., z_n = F_{n}(y_{n})\}$ and a uniform distribution could be considered. 

Here, we employ the Wasserstein-1 distance,
\begin{equation}\label{eq:MCB}
    \text{MCB} = \int_{0}^{1} \big|\hat{H}_z(u) - u \big|\ \mathrm{d} u,
\end{equation}
where $\hat{H}_z(u) = \tfrac{1}{n}\sum_{i = 1}^{n} \one\{ z_i \le u\}$, $u \in [0, 1]$, is the empirical distribution function of the $n$ PIT values. The larger MCB is, the more the distribution of the PIT values deviates from a standard uniform distribution, whereas smaller values of MCB indicate better calibration. The Wasserstein-1 distance can equivalently be written in terms of the quantile function associated with $\hat{H}_z$, and \eqref{eq:MCB} is estimated in practice by calculating the average absolute distance between the values $\{1/n, 2/n, \dots, 1\}$ and the sorted PIT values $\{z_{(1)}, \dots, z_{(n)}\}$ as $\tfrac{1}{n}\sum_{i=1}^{n}\lvert z_{(i)} - i/n \rvert$.

The MCB term of \eqref{eq:MCB} provides a measure of probabilistic miscalibration, and the regularized loss function of \eqref{eq:regu} therefore encourages forecast models to issue probabilistically calibrated forecasts. However, probabilistic calibration does not imply that the forecasts are calibrated in the tails. Optimizing a weighted scoring rule during model training can emphasize the tails, though the weighted scoring rule may reward discriminative ability more than calibration, meaning the resulting forecasts will not necessarily be tail calibrated. If we wish to enforce tail calibration, then we could adapt \eqref{eq:regu} by replacing the measure of miscalibration (MCB) with a measure of tail miscalibration (TMCB),
\begin{equation}\label{eq:tregu}
\frac{1}{n} \sum_{i=1}^{n} S(F_{i}, y_{i}) + \gamma \text{TMCB},
\end{equation}
where $\gamma \geq 0$. The measure of tail miscalibration again depends on $\{(F_1, y_1), \dots, (F_n, y_n)\}$, as well as the threshold of interest.

We employ a similar measure to quantify tail miscalibration:
\begin{equation}\label{eq:TMCB}
\text{TMCB} = \int_0^1 \big| \hat{Q}_{t}(u) - u \big|\ \mathrm{d}u,
\end{equation}
where $\hat{Q}_t$ is as defined in \eqref{eq:decomposition}. TMCB tends towards MCB as the threshold $t \to -\infty$. To estimate \eqref{eq:TMCB}, we use the average absolute difference $\tfrac{1}{n_t}\sum_{i=1}^{n_t}\lvert \hat{O}_t \cdot z_{(i), t} - i/n_t \rvert$, where $z_{(1), t} \le \dots \le z_{(n_t), t}$ are the order statistics of the empirical CPIT values, with $n_t = |\Ii_t|$. This exploits the decomposition $\hat{Q}_t(u) = \hat{O}_t\cdot\hat{H}_{z_t}^{-1}(u)$ in \eqref{eq:decomposition}. 

Instead of using $\hat{Q}_t$ to define TMCB, one could also quantify tail miscalibration by penalizing the deviation of the distribution of CPIT values from a standard uniform distribution, for example, by replacing $\hat{Q}_t(u)$ in \eqref{eq:TMCB} with the empirical distribution $\hat{H}_{z_t}(u)$. While this approach would similarly penalize forecasts with a miscalibrated excess distribution, it would neglect the forecast probability that an extreme event occurs. Hence, as we show in Appendix \ref{app:sec:penalizing-cpit}, this yields CPIT values that are closer to uniform, but severely deteriorates forecast performance with respect to all other metrics considered herein. This highlights the need to assess the calibration of both threshold exceedance probabilities (forecasts for extreme event occurrence) as well as the forecast excess distribution (forecasts for extreme event severity), and is similar in essence to the forecaster’s dilemma discussed by \cite{LerchEtAl2017}.

Divergences other than the Wasserstein-1 divergence could analogously be considered for the definition of MCB and TMCB, and we additionally studied the Wasserstein-2 and Kolmogorov-Smirnov distances. 
The results were qualitatively similar in all cases, though the Wasserstein distances were found to penalize more homogeneously across the domain of PIT and $\hat{Q}(u)$ values, thereby providing smoother objective functions for parameter optimization. The Cram{\'e}r distance could also be employed for this purpose. \cite{Wilks2018} demonstrated that the Cram{\'e}r distance often yields more powerful calibration tests than the Wasserstein-1 distance, though we conjecture that the results will be similar to those for the Wasserstein distance when used for optimization, since the gradients of these two divergence measures point in the same direction. The Wasserstein distances have the advantage that they can be calculated efficiently using the order statistics of the CPIT values. In the following, we therefore only present results for the Wasserstein-1 distance.

The miscalibration and tail miscalibration penalties do not correspond to average scores resulting from proper scoring rules. However, if the forecast is equal to the conditional distribution of the outcome given the covariates, then it will be both probabilistically calibrated and probabilistically tail calibrated. In this case, the (tail) miscalibration penalties will be equal to zero, their minimum value. As a consequence, if the model class is well-specified and the baseline scoring rule is strictly proper, then the regularized loss functions are still minimized uniquely at the true conditional distribution. That is, the regularization terms do not affect the consistency of the optimum score estimators.

\section{Simulation example}\label{sec:simstudy}

\begin{figure}[!h]
    \centering
    \includegraphics[width=0.9\linewidth]{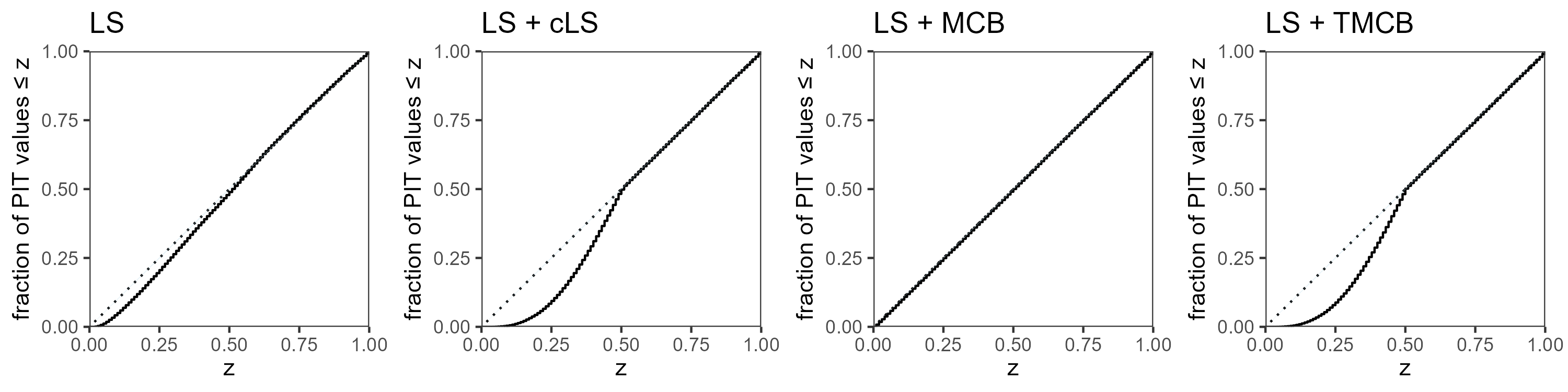}
    \includegraphics[width=0.9\linewidth]{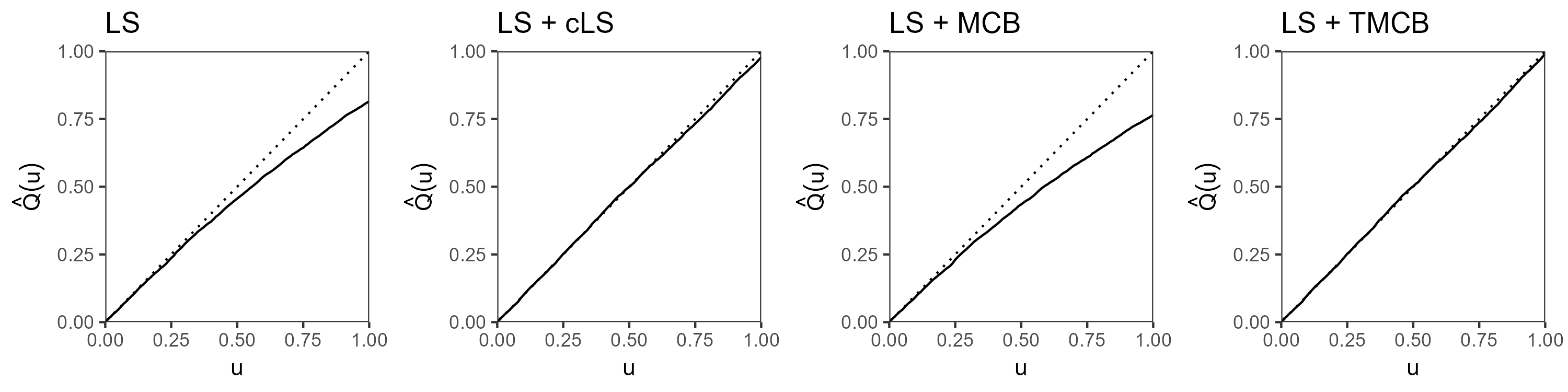}
    \caption{Calibration plots when estimating $a$ using standard maximum likelihood (first column), and when regularizing with the censored likelihood score (second), miscalibration (third), and tail miscalibration (fourth). Results are shown for $\gamma=2$ when assessing standard calibration (top row), and tail calibration (bottom).}
    \label{fig:ss2_cal}
\end{figure}

To demonstrate how the regularization terms described above affect estimated forecast model parameters, consider a simple example using simulated data. Suppose $Y \mid \mu \sim N(\mu, 1)$, where $\mu \sim \text{Unif}(-3, 3)$, and define a random variable $\tau$ that is equal to $1$ or $-1$ with equal probability, independently of $(Y, \mu)$. Suppose we are interested in outcomes $Y$ that exceed the threshold $t = 3.23$, roughly the 95th percentile of the unconditional distribution of $Y$.

Let $F_{1} = \frac{1}{2}N(\mu, 1) + \frac{1}{2}N(\mu + \tau, 1)$, and let $F_{2}$ be such that 
\[
F_{2}(x) = 
\begin{cases}
    \Phi \left( x - \mu \right) & \text{if $x \ge \mu$},\\
    \Phi \left( \frac{x - \mu}{2} \right) & \text{if $x < \mu$}.
\end{cases}
\]
The forecast $F_{1}$ is a slight adaptation of the unfocused forecaster of \citet[Example 3]{GneitingEtAl2007}, which is a probabilistically calibrated forecast for $Y$, but is not probabilistically tail calibrated. The forecast $F_{2}$ is equal to the true conditional distribution of $Y$ given $\mu$ for all $x \ge \mu$, but has the incorrect scale below $\mu$. Since $\mu < t$ almost surely, $F_2$ is therefore probabilistically tail calibrated but not probabilistically calibrated.

Suppose that we wish to estimate the optimal linear combination of the two forecasts $F_{1}$ and $F_{2}$. That is, we wish to estimate the optimal mixing parameter $a \in [0, 1]$ in the combined forecast distribution $F_{a} = a F_{1} + (1 - a)F_{2}$. We estimate $a$ by minimizing the log score (maximizing the likelihood) of the combined forecast $F_{a}$ over $n = 100,000$ realisations of $\mu$, $\tau$, and $Y$. The resulting parameter estimate is $\hat{a} = 0.72$, meaning $F_{1}$ is assigned a higher weight in the combination. The combined forecast $F_{\hat{a}}$ achieves an average log score of 1.52 over the $n$ realisations, compared with 1.53 and 1.58 for $F_{1}$ and $F_{2}$, respectively. However, while $F_{\hat{a}}$ is the optimal combination of $F_{1}$ and $F_{2}$ according to the log score, the resulting forecasts are neither calibrated nor tail calibrated. Calibration plots are displayed in Figure \ref{fig:ss2_cal}. 

\begin{figure}[!h]
    \centering
    \includegraphics[width=0.58\linewidth]{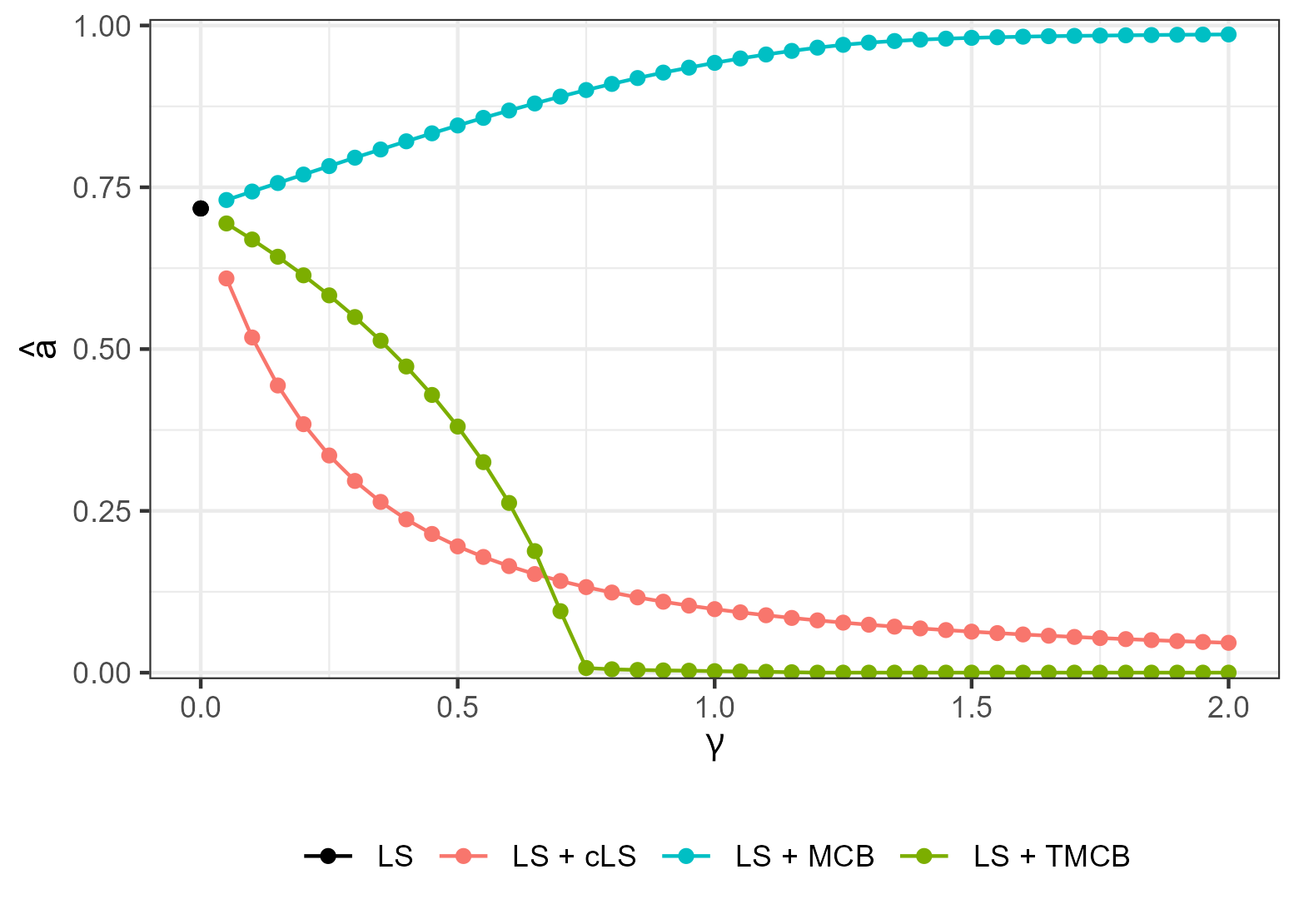}
    \caption{Estimated parameter $a$ in the simulation example as a function of $\gamma$ when penalizing miscalibration (blue), tail miscalibration (green), and censored likelihood score (red).} 
    \label{fig:ss2_a_gam}
\end{figure}

Consider now the parameters estimated when we minimize the log score with the regularization terms considered in the previous section. Regularization is performed using a penalty for the censored likelihood score (cLS), overall miscalibration (MCB), and tail miscalibration (TMCB). The cLS and TMCB terms are calculated using the threshold $t$. The estimated parameters for all methods are shown as a function of $\gamma$ in Figure \ref{fig:ss2_a_gam}. When penalizing overall miscalibration during model training, the estimated mixing parameter tends to one as $\gamma$ increases, recovering the unfocused forecaster $F_1$. The average log score of $F_{1}$ is marginally worse than that of $F_{\hat{a}}$, but $F_{1}$ is probabilistically calibrated whereas $F_{\hat{a}}$ is not (Figure \ref{fig:ss2_cal}), confirming that the miscalibration penalty yields forecasts that are better calibrated.

By instead penalizing tail miscalibration during model training, the estimated mixing parameter tends to zero as $\gamma$ increases, recovering the piecewise forecast $F_{2}$. The average log score of $F_{2}$ is also worse than that of $F_{\hat{a}}$, but $F_{2}$ is probabilistically tail calibrated whereas $F_{\hat{a}}$ is not (Figure \ref{fig:ss2_cal}), confirming that the tail miscalibration penalty yields forecasts that are better tail calibrated. 

Similar results are observed when regularizing with the censored likelihood score during model training. The censored likelihood score emphasizes outcomes above the threshold $t$ when calculating forecast accuracy, which implicitly encourages tail calibration. The estimated mixing parameter therefore again tends towards zero as $\gamma$ increases. The parameter tends to zero at a slower rate than for tail miscalibration regularization, but since the censored likelihood score is generally on a different scale to the measures of miscalibration, the dependence on $\gamma$ is not directly comparable between these methods; the $\gamma$ values in Figure \ref{fig:ss2_a_gam} for this method have been rescaled by a constant factor to aid visual comparison. 

\section{Weather forecasting application}\label{sec:applications}

Consider now an application of the regularization approaches in the context of weather forecasting. We apply the approaches to a hierarchy of increasingly complex statistical post-processing models, which issue probabilistic weather forecasts based on the output of numerical weather prediction models; further details about statistical post-processing can be found in \cite{VannitsemEtAl2018}. The three statistical models we consider are: ensemble model output statistics (EMOS), a simple linear parametric approach that is well established in operational post-processing suites \citep{GneitingEtAl2005}; distributional regression networks (DRNs), a non-linear extension of EMOS models based on neural networks \citep{rasp_neural_2018}; and conditional generative models (CGMs), a non-parametric generative model-based framework with a neural network architecture \citep{chen_generative_2024}. These methods are state-of-the-art in the post-processing literature, and we study how adapting the loss function used to train these models can influence the behavior of the resulting forecasts, particularly when interest is on extreme events.


\subsection{Data and methods}

\begin{figure}[!http]
    \centering
    \includegraphics[width=0.4\linewidth]{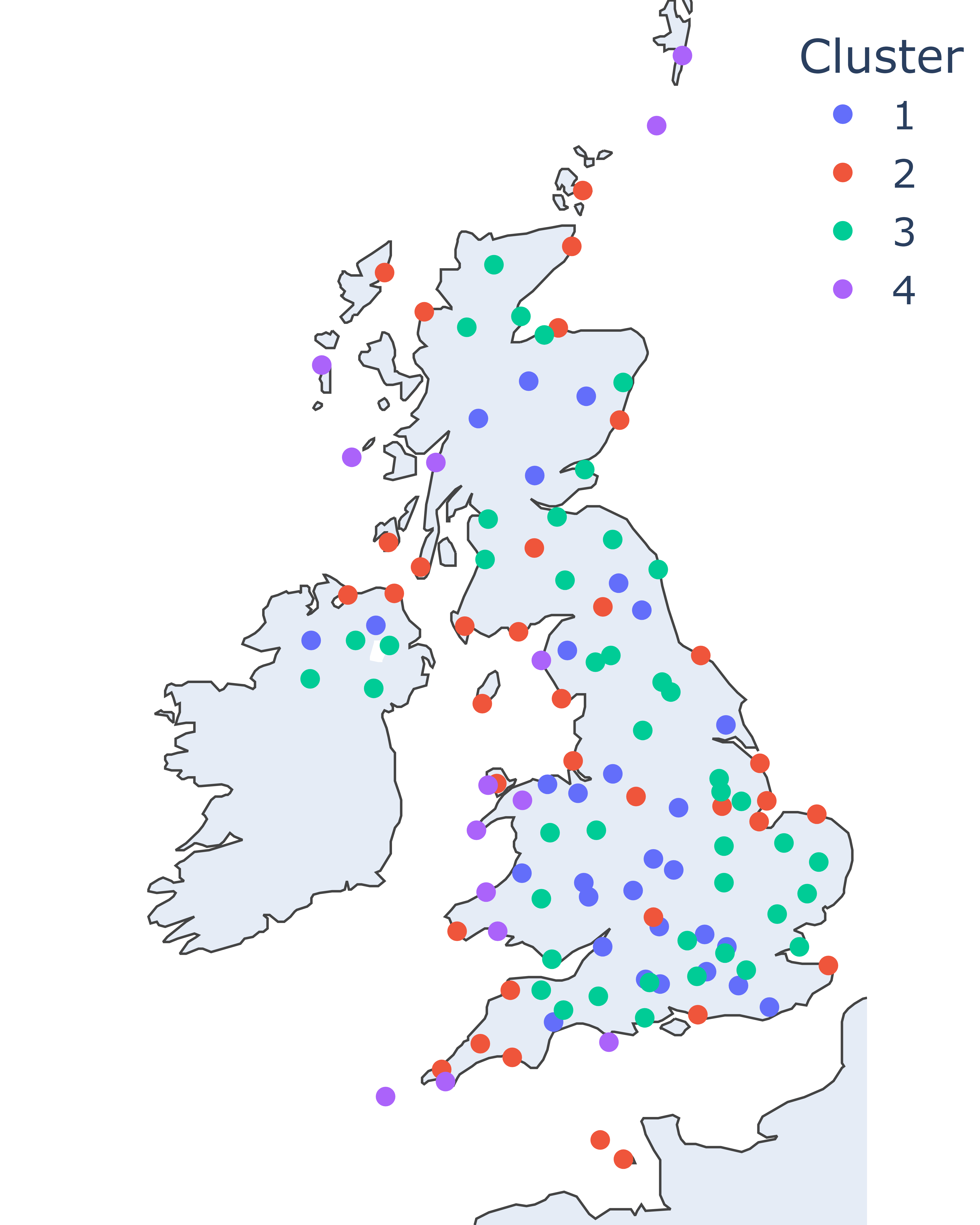}
    \caption{Locations of the 124 weather stations considered in the application. The color of each point corresponds to the cluster it is assigned to for the semi-local parameter estimation of the EMOS model (see Section \ref{sec:EMOS}). }
    \label{fig:locations}
\end{figure}

We consider forecasts of 10m wind speed from the UK Met Office's global ensemble prediction system \citep[MOGREPS-G;][]{walters_met_2017, porson_recent_2020}. The gridded MOGREPS-G output is bilinearly interpolated to the locations of 124 synoptic observation stations, shown in Figure \ref{fig:locations}, at which wind speed observations are available. We consider forecasts initialized at 00UTC and issued 72 hours in advance. Forecasts and observations between 1 April 2019 and 31 December 2020 are used to train the statistical post-processing methods, which are then evaluated on data between 1 January 2021 and 31 March 2022. 

The three post-processing models are discussed in the following sections. All models are trained by optimizing the CRPS, which is the most popular scoring rule to assess probabilistic weather forecasts. The same models are then additionally trained 
using the CRPS with the miscalibration, tail miscalibration, and twCRPS regularization terms discussed in Section \ref{sec:enforce}, with the results then compared to the baseline CRPS-trained models with no regularization. We explore the sensitivity of the results to different values of the penalty parameter $\gamma$.

All models use the same covariates: the mean and standard deviation of the MOGREPS-G ensemble forecast at the corresponding time and location, as well as the sine and cosine of the day of the year, allowing the model to capture seasonal variations in the wind speed; these harmonic terms additionally ensure that model performance is more consistent across the year, helping to account for climatological differences in the training and test (evaluation) datasets that arise due to them containing data for differing months. Further details about the model design and implementation are provided in Appendix \ref{sec:app_model}.

For the definition of an extreme event, we focus on an absolute wind speed threshold of 12.5m/s which is roughly equal to the 97.5th percentile of the distribution of wind speeds in the training data (across all stations). This threshold is used for both model training and evaluation. A range of different thresholds were also studied, all of which drew similar conclusions.

\subsection{Ensemble Model Output Statistics} \label{sec:EMOS}

\begin{figure}[!htb]
    \centering
    \includegraphics[width=0.9\linewidth]{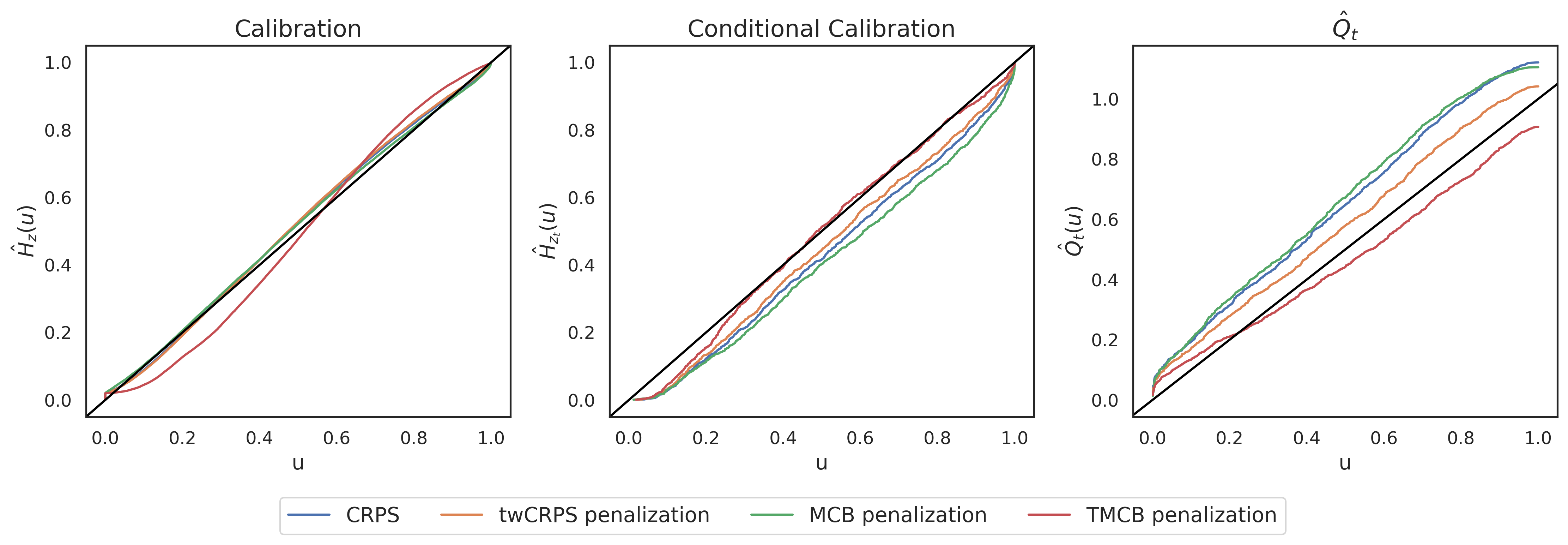}
    \caption{Calibration plots for EMOS models trained by minimizing the CRPS with different regularization terms, for $\gamma = 5$. Left: empirical distribution function of PIT values. Middle: empirical distribution function of conditional PIT values. Right: $\hat{Q}_t(u)$ as a function of $u \in [0, 1]$. In all cases, a calibrated forecast should result in lines close to the diagonal line shown in black. Results are pooled across all stations.}
    \label{fig:EMOS_1}
\end{figure}

Ensemble model output statistics \citep[EMOS;][]{jewson_new_2004, GneitingEtAl2005} is a popular post-processing technique that models the target variable using a parametric distribution whose location and scale parameters depend linearly on covariates. This can be thought of as a particular Generalized Additive Model for Location, Scale and Shape \citep[GAMLSS;][]{RigbyEtAl2005}. Following \cite{thorarinsdottir_probabilistic_2010}, we model wind speed using a truncated normal distribution. For parameter optimization, we use the closed-form expressions of the CRPS and twCRPS for a truncated normal distribution given by \cite{thorarinsdottir_probabilistic_2010} and \cite{wessel_lead-time-continuous_2024}, respectively.


EMOS parameters are estimated separately for four clusters of the 124 weather stations. The clusters are determined based on the distribution of wind speed observations in the training dataset at each location. This corresponds to a semi-local post-processing approach, which has been proposed as a means to augment training data \citep[see e.g.][]{lerch_similarity-based_2016}, making it particularly useful when interest is on extreme events. The resulting clusters are shown in Figure \ref{fig:locations}. Further details about the model design, optimization framework, and the semi-local estimation approach are given in Appendix \ref{sec:app_EMOS}. To ensure consistency with the evaluation of the other models in the following sections, the tail calibration of the EMOS models in each cluster is assessed using the same threshold of 12.5m/s. However, we additionally analyzed the results for cluster-wise thresholds, and the findings were analogous (not shown).

Figure \ref{fig:EMOS_1} presents calibration plots for the forecasts issued by EMOS models trained using the CRPS with different regularization terms (with $\gamma = 5.0$). All models issue reasonably well calibrated forecasts, with the distribution function of empirical PIT values lying close to the diagonal, though penalizing tail miscalibration results in forecasts that are slightly overdispersed. As expected, the CRPS-trained model yields forecasts that appear calibrated according to the standard notion of probabilistic calibration. However, these forecasts appear miscalibrated when predicting more extreme wind speeds: the conditional PIT values suggest that the estimated tail is too light. This reinforces the idea that standard checks for probabilistic calibration are not sufficient to infer whether or not the forecasts are tail calibrated.

Penalizing overall miscalibration during model training has little impact on the resulting forecasts in this case, and these forecasts are therefore similarly miscalibrated in the tails. In contrast, regularizing the loss function using the twCRPS or the measure of tail miscalibration leads to more reliable forecasts for extreme events. This can be seen both from improvements in the uniformity of CPIT values, as well as in $\hat{Q}_t$ values that are closer to the diagonal. However, this comes at the expense of standard calibration.

\renewcommand{\arraystretch}{1.2}
\begin{table}
\centering
\begin{tabular}{ll|ccc|}
\cline{3-5}
                                                     &                   & \multicolumn{3}{c|}{Regularization term}                                                     \\ \cline{3-5} 
                                                     &                   & \multicolumn{1}{c|}{MCB}           & \multicolumn{1}{c|}{TMCB}           & twCRPS        \\ \hline
\multicolumn{1}{|l|}{\multirow{2}{*}{Proper scores}} & CRPS skill (\%)   & \multicolumn{1}{c|}{-0.28}         & \multicolumn{1}{c|}{-4.77}          & -0.09         \\
\multicolumn{1}{|l|}{}                               & twCRPS skill (\%) & \multicolumn{1}{c|}{-1.04}         & \multicolumn{1}{c|}{-1.16}          & \textbf{0.12} \\ \hline
\multicolumn{1}{|l|}{\multirow{2}{*}{Calibration}}   & MCB skill (\%)    & \multicolumn{1}{c|}{\textbf{17.1}} & \multicolumn{1}{c|}{-187.19}        & -13.55        \\
\multicolumn{1}{|l|}{}                               & TMCB skill (\%)   & \multicolumn{1}{c|}{-10.46}        & \multicolumn{1}{c|}{\textbf{65.36}} & 44.81         \\ \hline
\end{tabular}
\caption{Relative improvement in evaluation metrics for the regularized EMOS models compared to the baseline CRPS-trained EMOS model. Results are shown for $\gamma = 5.0$. The largest improvement for each metric is displayed in bold. The average CRPS of the baseline model is 1.01.}
\label{tab:EMOS_1}
\end{table}

\begin{figure}[t!]
    \centering
    \includegraphics[width=0.5\linewidth]{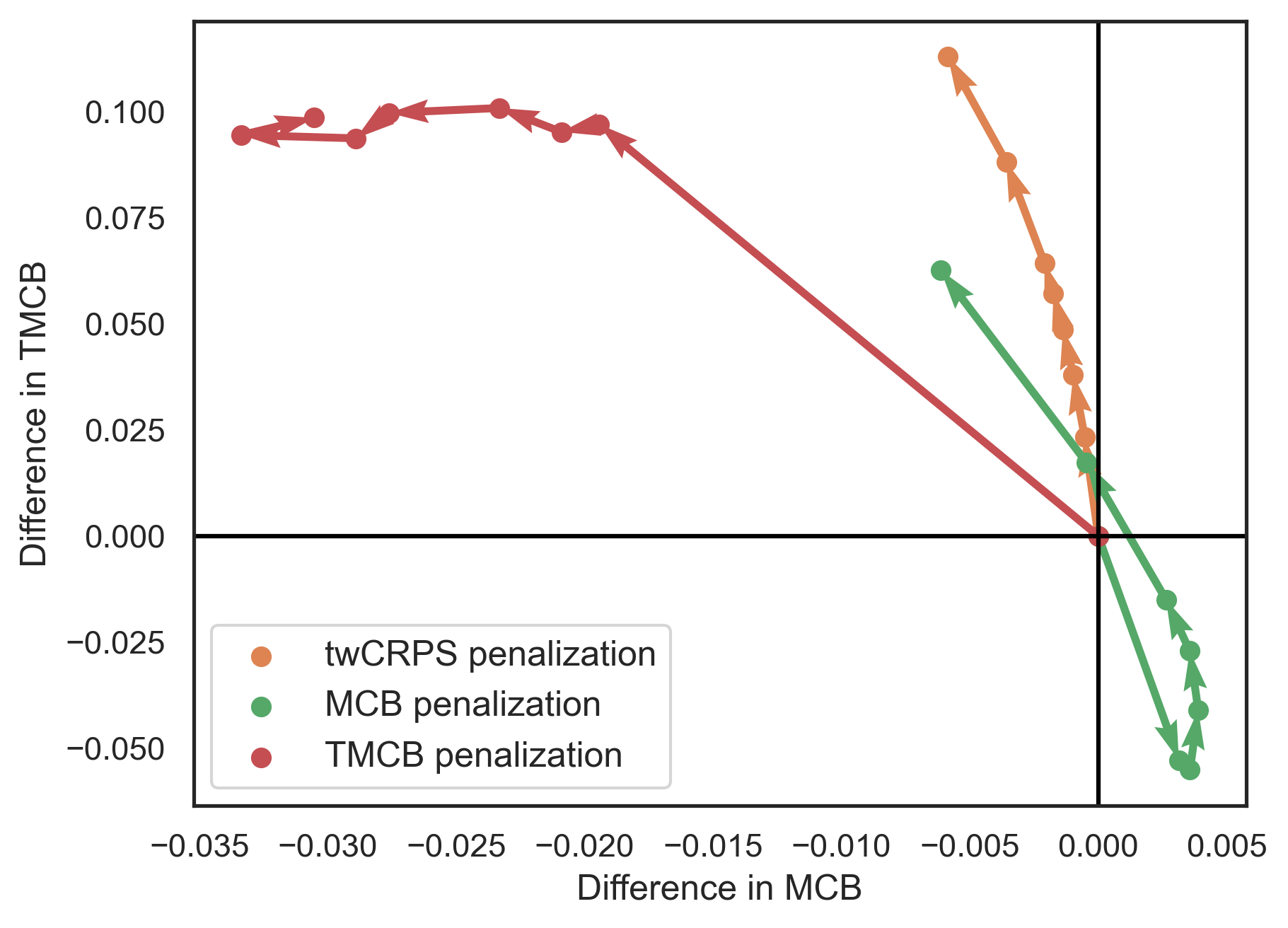}
    \caption{Absolute difference in MCB versus the absolute difference in TMCB for the EMOS models trained using the different regularization terms, relative to the CRPS-trained EMOS model with no regularization. Positive values indicate better (T)MCB of the regularized model compared to the CRPS-trained baseline. Results are shown for $\gamma = 0$ (the point at (0, 0)) and $\gamma \in \{1, 2, 3, 4, 5, 10, 20\}$, with arrows denoting the transitions between consecutive $\gamma$ values. Note the different scales of the x- and y-axes.}
    \label{fig:EMOS_2}
\end{figure}

Table \ref{tab:EMOS_1} quantifies the improvement of the various regularization methods (again with $\gamma = 5.0$) relative to the CRPS-trained model when assessed using the CRPS, twCRPS, and the different measures of calibration. Penalizing tail miscalibration during model training can improve the tail calibration of the resulting forecasts by more than 60\%, though the overall calibration becomes 187\% worse; it should be noted that since the baseline CRPS-trained model is reasonably well calibrated overall but not in the tails, the 60\% improvement in tail calibration corresponds to a larger absolute difference than the 187\% deterioration in overall calibration (see Figure \ref{fig:EMOS_2}). Adding the twCRPS to the loss function results in similar but more moderate changes to the considered metrics, with a minor improvement in twCRPS relative to the baseline.

Figure \ref{fig:EMOS_2} presents the absolute difference in overall miscalibration and tail miscalibration from the baseline CRPS-trained model as a function of $\gamma$. Since the twCRPS is generally on a smaller scale than the measures of miscalibration, it is difficult to directly compare results for the different methods for a specific value of $\gamma$. When penalizing the twCRPS or tail miscalibration, increasing $\gamma$ results in a larger improvement with respect to tail miscalibration (before plateauing for larger $\gamma$), and a larger deterioration in overall calibration. Penalizing overall miscalibration generally has small effects on both overall as well as tail calibration. For larger values of $\gamma$, no improvement in overall calibration can be seen on the test dataset, though such an improvement exists on the training dataset. These results need to be taken with care, as estimation of the EMOS parameters with (T)MCB regularization can become somewhat unstable for large values of $\gamma$. 

\subsection{Distributional Regression Networks} \label{sec:DRN}

\begin{figure}[h!]
    \centering
    \includegraphics[width=0.9\linewidth]{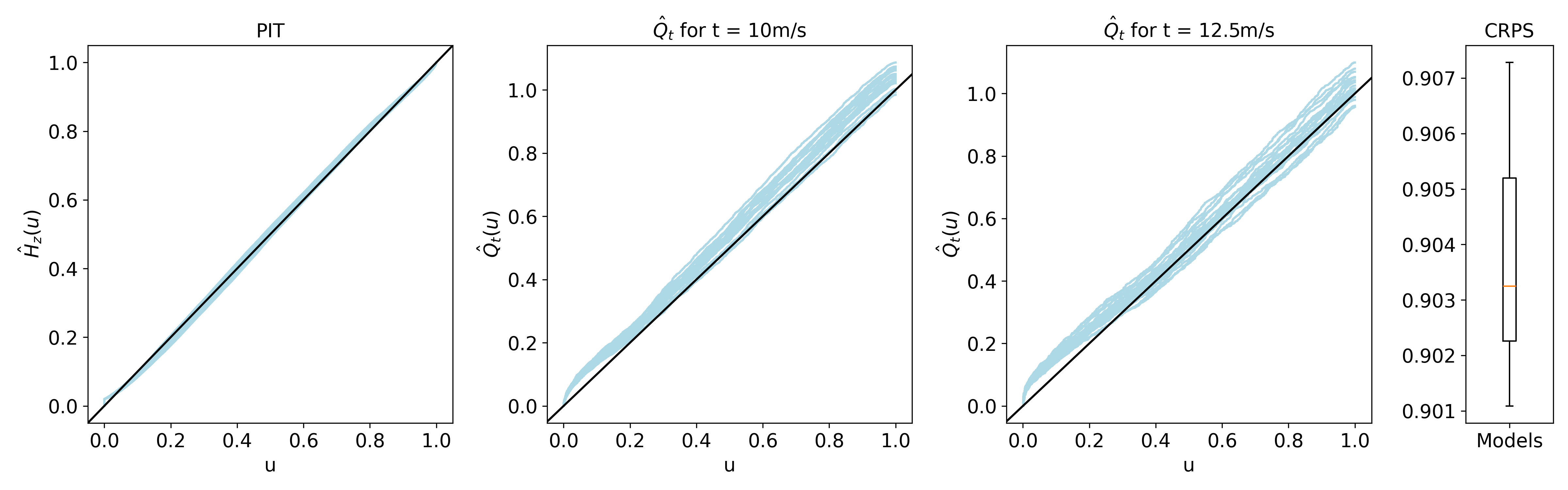}
    \caption{Empirical distribution function of PIT values (first panel) and tail calibration diagnostic plots (second and third panels) for the 100 CRPS-trained DRN models, each shown by a light blue line. Tail calibration plots are shown at two thresholds, $t = 10$m/s and $t = 12.5$m/s. Perfect calibration is denoted by the diagonal black lines. The final panel shows the distribution of the CRPS for the 100 DRN models.}
    \label{fig:DRN_1}
\end{figure}

Distributional regression networks \citep[DRNs,][]{rasp_neural_2018} can be seen as a non-linear extension of EMOS. Wind speed is again assumed to follow a truncated normal distribution, but the location and scale of the distribution are now modeled as the output of a feedforward neural network with the covariates as inputs. Following \cite{rasp_neural_2018}, the station index is also incorporated as a covariate via a learned embedding layer, allowing one spatially-aware DRN model to be fit using the data from all stations. More details on training the DRNs are given in Appendix \ref{sec:app_DRN}.

Since DRNs output a truncated normal predictive distribution, the network parameters can again be trained using available closed-form expressions of the CRPS and twCRPS. However, in contrast to EMOS models, DRNs are heavily over-parametrized, and (CRPS-based) neural network training presents a non-convex optimization problem. Hence, when training multiple DRNs, the resulting neural networks will generally differ in their parameters, due to randomness in both the initial weight values and the stochastic gradient descent algorithm. To circumvent this, we fit the DRN 100 times, and analyse the resulting scores and calibration metrics across these 100 models; this allows us to understand how the results vary for different learned parameter configurations. This is repeated when training the DRN using each regularization term.

Figure \ref{fig:DRN_1} shows the calibration and tail calibration of the 100 DRN models trained by minimizing the CRPS without regularization, as well as the resulting CRPS in the test data. There is low variation in the CRPS of the 100 DRN models, and in the distribution of the corresponding PIT values, with all models yielding well calibrated forecasts. However, the resulting forecasts can behave very differently with respect to tail calibration, despite the fairly rigid parametric assumptions made by the DRN. Some models yield forecasts that appear reasonably well calibrated in the tails, while others exhibit severe miscalibration. Both the CRPS and the overall calibration are insensitive to this variation in tail calibration. This variation stems solely from training randomness, highlighting the need for strategies to reduce and better control such variability.

\begin{figure}[!t]
    \centering
    \includegraphics[width=\linewidth]{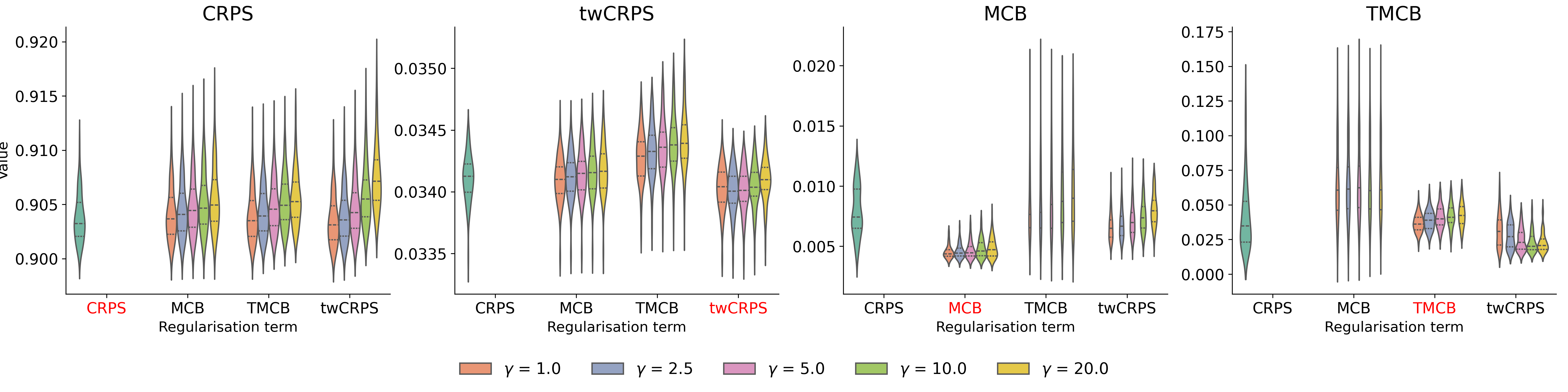}
    \caption{Violin plots of metrics for the 100 DRN models trained using different regularization terms. Different panels correspond to different evaluation metrics, while the x-axes labels correspond to the different regularization terms. Red x-axis labels correspond to when the evaluation metric is the same as the metric used for regularization.}
    \label{fig:DRN_2}
\end{figure}

We finetune the CRPS-trained DRN models using the different regularization terms discussed above (for details, see Appendix \ref{sec:app_DRN}). Figure \ref{fig:DRN_2} shows the distribution of the CRPS, twCRPS, and measures of miscalibration for the 100 DRN models trained using the CRPS with the different regularization terms, for different values of $\gamma$. All models yield similarly accurate forecasts when assessed using the CRPS and twCRPS. Penalizing tail miscalibration or twCRPS during training appears not to strongly decrease the accuracy of the forecasts in this study, though it also generally does not improve the twCRPS. However, training with these penalties strongly reduces the variation in the tail miscalibration measures of the 100 DRN models. 

Table \ref{tab:DRN_1} shows the relative improvement in each metric compared to the CRPS-trained baseline, for $\gamma = 5.0$. The improvement is based on the average value of the evaluation metric calculated over the 100 DRN models. The results are qualitatively similar to those obtained for the EMOS forecasts. Penalizing overall miscalibration improves the calibration of the resulting forecasts, at the expense of accuracy and tail calibration. Conversely, penalizing tail miscalibration during model training improves tail calibration at the expense of overall calibration. Adding the twCRPS to the loss function yields yet larger improvements in the tail calibration of the DRN models, as well as a slight improvement in overall calibration.

\begin{figure}[b!]
    \centering
    \includegraphics[width=0.5\linewidth]{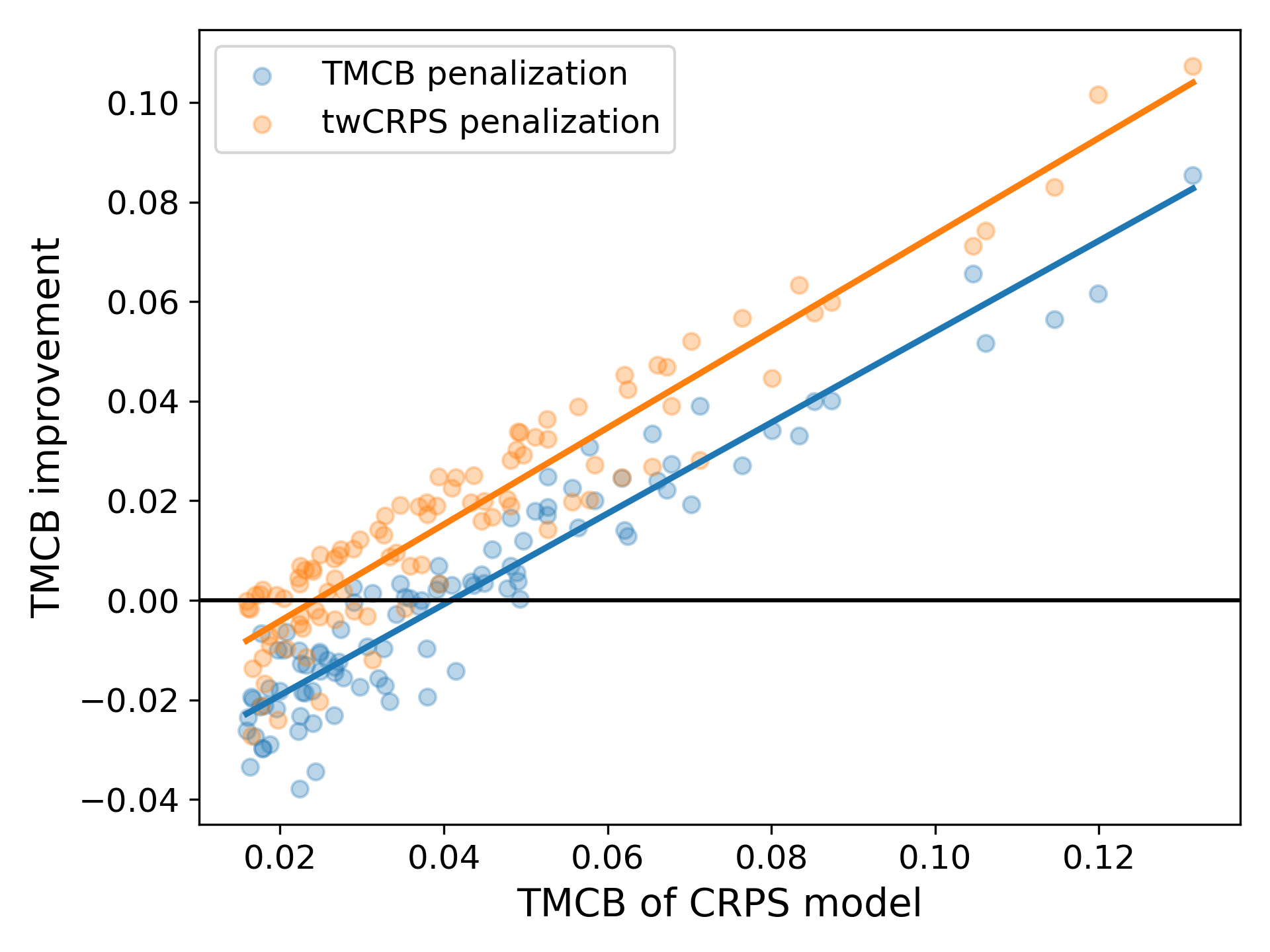}
    \caption{Improvement in TMCB obtained from the regularized model ($\gamma = 5.0$) relative to the baseline CRPS-trained model, plotted against the TMCB of the baseline model. Results are shown for twCRPS (orange) and tail miscalibration (blue) regularization. Each point corresponds to one of the 100 DRN models, with the solid lines representing linear regression fits to these 100 points. }
    \label{fig:DRN_3}
\end{figure}

The improvement in tail calibration is not consistent across the 100 DRN models, and depends on the miscalibration of the CRPS-trained model. Figure \ref{fig:DRN_3} displays the (absolute) improvement in tail calibration against the tail miscalibration of the CRPS-trained model. Results are shown for the models trained using regularization with the twCRPS and the tail miscalibration, for $\gamma = 5.0$. Intuitively, the larger the tail miscalibration of the CRPS-trained model, the larger the improvements obtained by penalizing tail miscalibration during model training. When the CRPS-trained model already issues reasonably well calibrated forecasts in the tails, the additional regularization term can introduce unnecessary uncertainty into the parameter estimation, which can slightly hinder the tail calibration of the resulting forecasts.

\renewcommand{\arraystretch}{1.2}
\begin{table}
\centering
\begin{tabular}{ll|ccc|}
\cline{3-5}
                                                     &                   & \multicolumn{3}{c|}{Regularization term}                                                     \\ \cline{3-5} 
                                                     &                   & \multicolumn{1}{c|}{MCB}            & \multicolumn{1}{c|}{TMCB}          & twCRPS        \\ \hline
\multicolumn{1}{|l|}{\multirow{2}{*}{Proper scores}} & CRPS skill (\%)   & \multicolumn{1}{c|}{-0.12}          & \multicolumn{1}{c|}{-0.12}         & -0.11        \\
\multicolumn{1}{|l|}{}                               & twCRPS skill (\%) & \multicolumn{1}{c|}{-0.11}          & \multicolumn{1}{c|}{-0.73}         & \textbf{0.28} \\ \hline
\multicolumn{1}{|l|}{\multirow{2}{*}{Calibration}}   & MCB skill (\%)    & \multicolumn{1}{c|}{\textbf{40.97}} & \multicolumn{1}{c|}{-15.83}        & 10.17         \\
\multicolumn{1}{|l|}{}                               & TMCB skill (\%)   & \multicolumn{1}{c|}{-59.08}         & \multicolumn{1}{c|}{2.33} & \textbf{48.90}         \\ \hline
\end{tabular}
\caption{Relative improvement of average evaluation metrics for the 100 regularized DRN models compared to the 100 baseline CRPS-trained DRN models. Results are shown for $\gamma = 5.0$. The largest improvement for each metric is displayed in bold.}
\label{tab:DRN_1}
\end{table}

\subsection{Conditional Generative Models}

\begin{figure}[h!]
    \centering
    \includegraphics[width=0.9\linewidth]{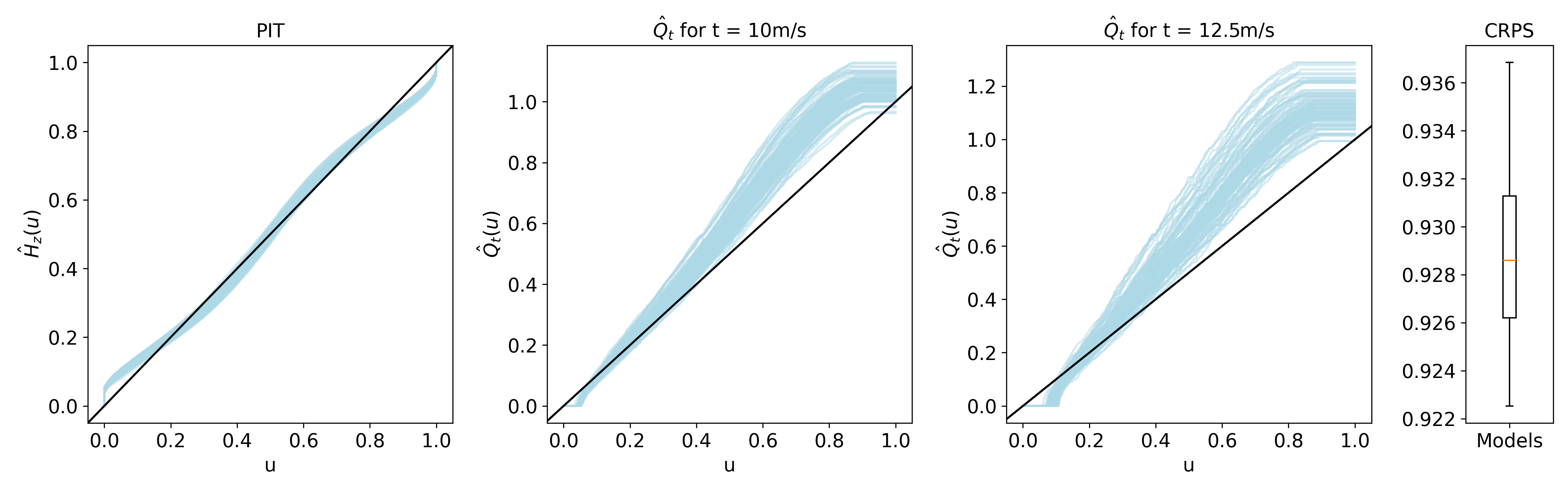}
    \caption{Empirical distribution function of PIT values (first panel) and tail calibration diagnostic plots (second and third panels) for the 100 CRPS-trained CGMs, each shown by a light blue line. Tail calibration plots are shown at two thresholds, $t = 10$m/s and $t = 12.5$m/s. Perfect calibration is denoted by the diagonal black lines. The final panel shows the distribution of the CRPS for the 100 CGMs.}
    \label{fig:CGM_1}
\end{figure}

Conditional generative models \citep[CGMs,][]{ chen_generative_2024} provide a non-parametric forecasting framework that constructs predictive distributions by learning a map between the covariates, a latent random variable, and the target variable. The map is typically modeled using a neural network. CGMs do not rely on parametric assumptions, giving them more flexibility than EMOS and DRN models. However, while EMOS and DRNs output a continuous parametric distribution, the CGM predictive distribution is only available via a sample of possible outcomes. We refer to \cite{chen_generative_2024, pacchiardi_probabilistic_2024, SchillingerEtAl2025} for more background on CGMs. 

As for the DRN, we fit one CGM to data from all stations, using an embedding layer to encode spatial information. This is repeated 100 times to account for uncertainty in the model training. The CGM model architecture is similar to \cite{chen_generative_2024}, except that while they designed CGMs to obtain multivariate predictive distributions, we implement them here in the univariate setting; that is, the map that we learn using the CGM is a function from the covariates and latent variable to the univariate wind speed at a given time and location. We sample 250 values from the generative model, resulting in predictive distributions that are reasonably smooth. Parameter estimation is performed using sample-based estimates of the CRPS and twCRPS, and an additional smoothing of the PIT and CPIT values is also applied during model training to ensure that the gradients of the miscalibration penalties exist. This is discussed in Appendix \ref{app:sec:penalizing-cpit}. 

Figure \ref{fig:CGM_1} shows calibration plots for predictive distributions obtained from the 100 CGM models, trained using the CRPS. The models show very little variability in overall calibration and CRPS. Minor deviation of the PIT values from the diagonal suggests that the CGM forecasts are not perfectly calibrated, while the forecasts are generally more accurate than the EMOS forecasts but less accurate than the DRN forecasts, as assessed using the average CRPS. There is again large variability in the tail calibration of the CRPS-trained CGM forecasts, though all 100 models yield miscalibrated forecasts for extreme outcomes.

\begin{figure}[t!]
    \centering
    \includegraphics[width=\linewidth]{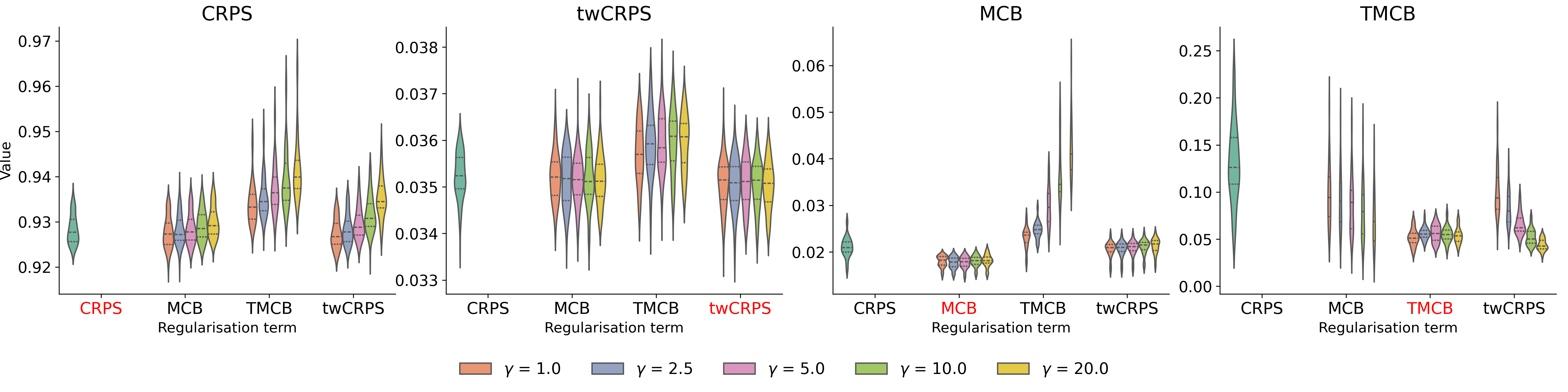}
    \caption{Violin plots of metrics for the 100 CGMs trained using different regularization terms. Different panels correspond to different evaluation metrics, while the x-axis labels correspond to the different regularization terms. Red x-axis labels correspond to when the evaluation metric is the same as the metric used for regularization.} 
    \label{fig:CGM_2}
\end{figure}

\begin{figure}[t!]
    \centering
    \includegraphics[width=0.5\linewidth]{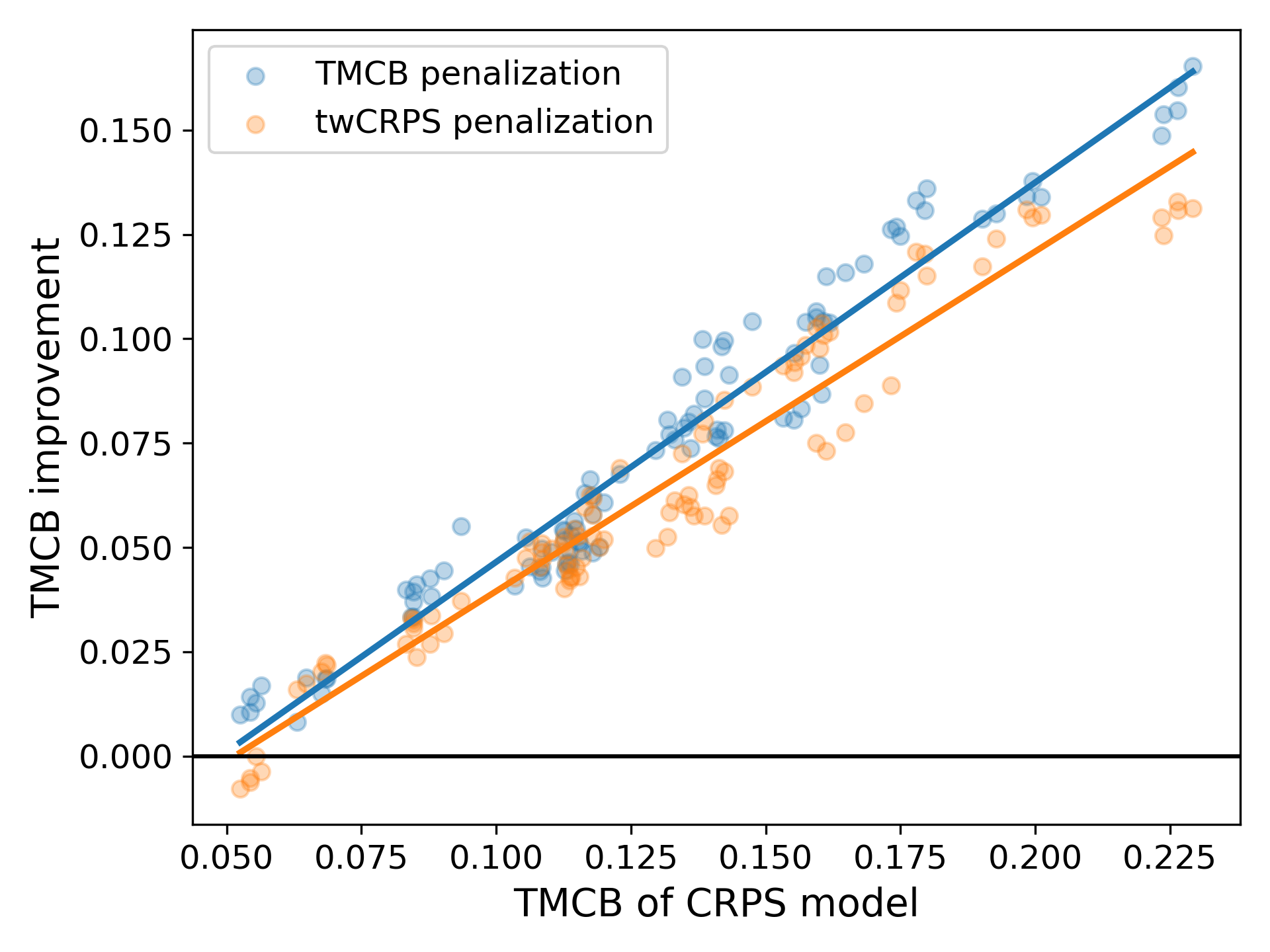}
    \caption{Improvement in TMCB obtained from the regularized model ($\gamma = 5.0$) relative to the baseline CRPS-trained model, plotted against the TMCB of the baseline model. Results are shown for twCRPS (orange) and tail miscalibration (blue) regularization. Each point corresponds to one of the 100 CGM models, with the solid lines representing linear regression fits to these 100 points.}
    \label{fig:CGM_3}
\end{figure}

We again finetune the CRPS-trained CGM models using the tail calibration penalties. Figure \ref{fig:CGM_2} shows the distribution of the evaluation metrics for the 100 CGM models trained using the CRPS with the different regularization terms, for different values of $\gamma$. Table \ref{tab:CGM_1} presents the relative improvement in average metrics compared with the baseline CRPS-trained CGMs, for $\gamma = 5.0$. The results are qualitatively similar to those obtained for EMOS and the DRNs: penalizing twCRPS or tail miscalibration during model training results in slightly worse CRPS and overall calibration, but can significantly improve the tail calibration of the resulting forecasts; the variation in tail miscalibration across the 100 models is also significantly decreased via these regularization terms; twCRPS regularization has less impact on the CRPS and overall calibration, but generally improves tail calibration less than the tail calibration regularization. 
Penalizing overall miscalibration during model training helps to address the slight miscalibration of the original CGM forecasts, while also slightly improving twCRPS and tail calibration, likely due to the increased spread of the resulting predictive distributions meaning that fewer observations fall outside the range of the 250 samples from the model. The improvement in tail calibration again depends on the original miscalibration of the CRPS-trained CGM (Figure \ref{fig:CGM_3}).

\renewcommand{\arraystretch}{1.2}
\begin{table}
\centering
\begin{tabular}{ll|ccc|}
\cline{3-5}
                                                     &                   & \multicolumn{3}{c|}{Regularization term}                                                      \\ \cline{3-5} 
                                                     &                   & \multicolumn{1}{c|}{MCB}            & \multicolumn{1}{c|}{TMCB}           & twCRPS        \\ \hline
\multicolumn{1}{|l|}{\multirow{2}{*}{Proper scores}} & CRPS skill (\%)   & \multicolumn{1}{c|}{-0.02}          & \multicolumn{1}{c|}{-0.98}          & -0.15         \\
\multicolumn{1}{|l|}{}                               & twCRPS skill (\%) & \multicolumn{1}{c|}{0.26}           & \multicolumn{1}{c|}{-2.03}          & \textbf{0.43} \\ \hline
\multicolumn{1}{|l|}{\multirow{2}{*}{Calibration}}   & MCB skill (\%)    & \multicolumn{1}{c|}{\textbf{15.45}} & \multicolumn{1}{c|}{-42.35}         & 1.15          \\
\multicolumn{1}{|l|}{}                               & TMCB skill (\%)   & \multicolumn{1}{c|}{32.80}          & \multicolumn{1}{c|}{\textbf{56.88}} & 49.28         \\ \hline
\end{tabular}
\caption{Relative improvement of average evaluation metrics for the 100 regularized CGMs compared to the 100 baseline CRPS-trained GCMs. Results are shown for $\gamma = 5.0$. The largest improvement for each metric is displayed in bold.}
\label{tab:CGM_1}
\end{table}

\section{Conclusions}\label{sec:conclusion}

This work studies how the loss function used to train probabilistic forecast models can be adapted to target extreme events. While parameters of probabilistic models are typically estimated by optimizing a proper scoring rule over a set of training data, if the model class is not well-specified or only a small training dataset is available, then the choice of scoring rule will influence the behavior of the resulting forecasts. In particular, there is no guarantee that the predictions obtained from the model will be calibrated or tail calibrated. 

In an application to wind speed forecasting, we demonstrate that three state-of-the-art statistical and machine learning models fail to issue forecasts that are calibrated when forecasting extreme outcomes. This includes forecasts issued by simple parametric approaches, distributional regression networks, and conditional generative models. We therefore consider how the training of these methods could be adapted to improve the reliability of the resulting forecasts.

We consider two approaches. The first uses a weighted scoring rule to target extreme outcomes during model training, as studied by \cite{WesselEtAl2024}. The motivation is that if forecasts for extreme events are to be evaluated using weighted scoring rules, then it makes sense to deploy them when fitting the forecast model; the resulting forecasts should be more accurate when assessed using the scores with which they are trained. Alternatively, the (tail) miscalibration could be penalized more directly by adding a measure of (tail) miscalibration as a regularization term. This measure of miscalibration does not encompass the information content of the forecast, and can therefore result in a larger deterioration of the forecast accuracy, though it generally produces more reliable forecasts. We find that both training with a tail calibration penalty and a weighted score improve out-of-sample tail calibration. This can, however, come at the expense of overall probabilistic calibration and forecast skill.


We studied extreme wind speed events defined as exceedances of a moderately high threshold, chosen using relevant quantiles of historical observations. However, such threshold exceedances do not necessarily correspond to impactful events in reality. While it would be desirable to consider higher thresholds, such an analysis is prohibited by data availability, which remains a crucial challenge for characterizing and forecasting extreme events. The motivation behind this work was to demonstrate how the loss function can be adapted to obtain better-calibrated forecasts for particular outcomes. Future work should assess how the proposed regularization strategies perform as the amount of training data changes, which would, for example, help to understand the scalability and robustness of the results in data-rich settings.

Nonetheless, while we focus on extreme events, the notion of tail calibration readily extends to other regions of the outcome space \citep{AllenEtAl2023,MitchellWeale2023}. Local miscalibration within these regions could similarly be penalized during model training to obtain more reliable forecasts for these particular outcomes. This could be used to obtain more reliable forecasts for outcomes within a particular range, below a threshold of interest, or in disjoint regions of interest, for example. The framework could additionally be extended to multivariate probabilistic forecasts. However, the calibration of multivariate forecasts has received much less attention in the literature, making it difficult to define notions of multivariate tail calibration. On the other hand, \cite{AllenEtAl2023b} introduce weighted multivariate scoring rules, and it would be interesting to assess whether they can be employed during model training to yield more accurate and reliable forecasts for multivariate extremes.

Other models, beyond the three considered in Section \ref{sec:applications}, can similarly be regularized to penalize (tail) miscalibration. Forecasting methods that do not rely on the optimization of an objective function, such as $k$-nearest neighbours, cannot, since the regularization relies on having a parameter estimate that can be pushed towards a better-calibrated solution. This regularization approach is also not necessary for forecast methods that are guaranteed to be calibrated in the training data by construction, such as isotonic distributional regression \citep[IDR;][]{HenziEtAl2021} and binning or analogue methods \citep[Proposition 2.4]{AllenEtAl2025}. However, the regularization approach is also applicable to methods that rely on objective functions that are not constructed using proper scoring rules.

The (tail) miscalibration regularization term penalizes probabilistic (tail) miscalibration, which is the notion of forecast calibration most commonly assessed in practice. Probabilistic calibration is a fairly weak notion of calibration \citep{GneitingResin2023}, though it can therefore be interpreted as a minimum requirement that probabilistic forecasts should satisfy. While it would be more desirable to penalize a stronger notion of calibration, such as auto-calibration \citep{Tsyplakov2011,GneitingRanjan2013}, stronger notions are generally difficult to calculate efficiently (or at all), limiting their applicability within loss functions.

The methods used herein weakly enforce tail calibration by adding an additional regularization term to the loss function. Future work could consider methods to strongly enforce tail calibration. For example, conformal prediction has recently received much attention in the machine learning literature as a means to issue predictions with theoretical, out-of-sample calibration guarantees \citep[see][for textbook introductions]{VovkEtAl2022,AngelopoulosEtAl2024}. Conformal predictive systems are generally not constructed using optimum score estimators, but it may be possible to introduce conformal prediction strategies that are tailored to the prediction of extremes. 

Similarly, the idea of regularizing the loss function relies on the idea that there are two objective functions that are to be minimized. In this paper, we combine these loss functions and study how forecast behavior changes as the mix between these losses is changed. However, one could also treat this as a multi-objective optimization problem, in which the multiple loss functions are to be optimized simultaneously. Multi-objective optimization is already commonly used in fields such as hydrology and climate science; \cite{zerenner_multi-objective_2021}, for example, simultaneously optimize the CRPS and root mean squared error of corresponding point predictions. Instead, a proper scoring rule, a measure of forecast miscalibration, and a measure of tail miscalibration could be included as three objective functions that are to be minimized simultaneously. In the neural network setting, multiple gradient descent methods \citep[e.g.][]{fliege_steepest_2000, desideri_multiple-gradient_2012, mercier_stochastic_2018} could be used for this. 


\section*{Acknowledgements}
Jakob Wessel was supported during this work by an Engineering and Physical Sciences Research Council (EPSRC) Doctoral Training Partnership (DTP) under grant award number 2696930. Sam Allen gratefully acknowledges funding from the Vector Stiftung through the Young Investigator Group ``Artificial Intelligence for Probabilistic Weather Forecasting''. The authors thank Johanna Ziegel and two anonymous reviewers for their insightful comments on previous versions of the paper. 

\section*{Data and code availability}
Code to reproduce the results herein is available at \url{https://github.com/jakobwes/Enforcing-tail-calibration}. Unfortunately the NWP forecast and observation data used in the case study in Section \ref{sec:applications} are proprietary to the UK Met Office and cannot be shared publicly by the authors. Readers may request this data directly from UK Met Office.

\bibliographystyle{apalike}
\bibliography{biblio.bib} 

\appendix

\section{Tail miscalibration interpretation}\label{app:estimators}

Let $O_t = \Q(Y > t) / \E[1 - F(t)]$, for $t \in \R$, and let $H_{Z_t}$ be the conditional distribution function of $F_t(Y)$ given $Y > t$, with inverse $H_{Z_t}^{-1}$. From Definition \ref{def:ptc}, a forecast $F$ is probabilistically tail calibrated at threshold $t$ if $O_t = 1$ and $H_{Z_t}(u) = H_{Z_t}^{-1}(u) = u$, for all $u \in [0, 1]$. The forecast can be interpreted as having a tail that is too light if $O_t > 1$ and $H_{Z_t}(u) < u$ (or $H_{Z_t}^{-1}(u) > u$) for all $u \in (0, 1)$, and too heavy if $O_t < 1$ and $H_{Z_t}(u) > u$  (or $H_{Z_t}^{-1}(u) < u$) for all $u \in (0, 1)$. If $O_t < 1$ but $H_{Z_t}(u) > u$, then the forecast underpredicts the average exceedance probability at threshold $t$, but generally overpredicts the probability of higher threshold exceedances (the opposite is true if $O_t > 1$ but $H_{Z_t}(u) < u$). In this case, it is more difficult to make conclusions regarding whether the forecast exhibits a tail that is too light or too heavy.

\cite{AllenEtAl2024} note that a forecast is probabilistically tail calibrated if and only if $R_t(u) = O_t H_{Z_t}(u) = u$ for all $u \in [0, 1]$. This allows tail miscalibration to be assessed in practice via one plot of $u \mapsto R_t(u)$, rather than two separate plots of $t \mapsto O_t$ and $u \mapsto H_{Z_t}(u)$. However, as the authors remark, the resulting plot is more difficult to interpret, since errors in $O_t$ and $H_{Z_t}$ can cancel each other out in such a way that $R_t(u)$ is close (albeit not exactly equal) to $u$, despite biases in both components. For example, for a forecast that is too light tailed ($O_t > 1$ and $H_{Z_t}(u) < u$ for all $u \in (0, 1)$), we cannot say whether $R_t(u)$ should be less than or greater than $u$. In fact, the plot of $u \mapsto R_t(u)$ will often cross the diagonal line, and the forecast may appear better tail calibrated compared to a forecast with the same $H_{Z_t}(u)$ but with $O_t = 1$. It is therefore difficult to conclude from the resulting plot that the forecast has a tail that is too light. Analogous arguments hold for a forecast that is too heavy tailed. 

This also limits the applicability of this approach for the purpose of weakly enforcing tail calibration. A measure of tail miscalibration could be derived by measuring the distance between $R_t(u)$ and $u$, which could be used as a regularization term to penalize tail miscalibration during model training, as proposed in Section \ref{sec:penalty}. However, the cancelling out of the biases in $R_t$ can yield a loss function that is more difficult to optimize; this was therefore found to perform worse in practice than the approach based on $Q_t(u) = O_t H_{Z_t}^{-1}(u)$ outlined in Section \ref{sec:calibration}. 

The quantity $Q_t$ partly remedies the issues caused by the cancelling out of biases in $O_t$ and $H_{Z_t}$. For example, if the forecast is too light-tailed, then $O_t > 1$ and $H_{Z_t}^{-1}(u) > u$, and hence $Q_t(u) > u$. Similarly, if the forecast is too heavy tailed, then $O_t < 1$, $H_{Z_t}^{-1}(u) < u$, and $Q_t(u) < u$. In this case, a larger bias in $O_t$ or $H_{Z_t}$ leads to a larger bias in $Q_t$. The resulting plot of $u \mapsto Q_t(u)$ is therefore much more interpretable than a plot of $u \mapsto R_t(u)$, since the biases in the individual components scale in the same direction. Of course, if $O_t > 1$ but $H_{Z_t}^{-1}(u) < u$, then $Q_t(u)$ may still be close to $u$ despite the biases in the individual components (which would not be the case for $R_t(u)$), but this combination is less common in practice, and the interpretation in this case is unclear anyway. 

Similar arguments hold when $O_t, H_{Z_t}$, and $ Q_t$  are replaced with the finite-sample estimates $\hat{O}_t, \hat{H}_{z_t}$, and $\hat{Q}_t$ in Equation \ref{eq:decomposition}, and $R_t$ is replaced by 
\[
    \hat{R}_{t}(u) = \frac{\sum_{i \in \Ii_{t}} \one\{z_{i,t} \leq u \}}{ \sum_{i = 1}^{n} [1 - F_{i}(t)]}, \quad u \in [0, 1].
\]

\begin{figure}[h]
    \centering
    \includegraphics[width=\linewidth]{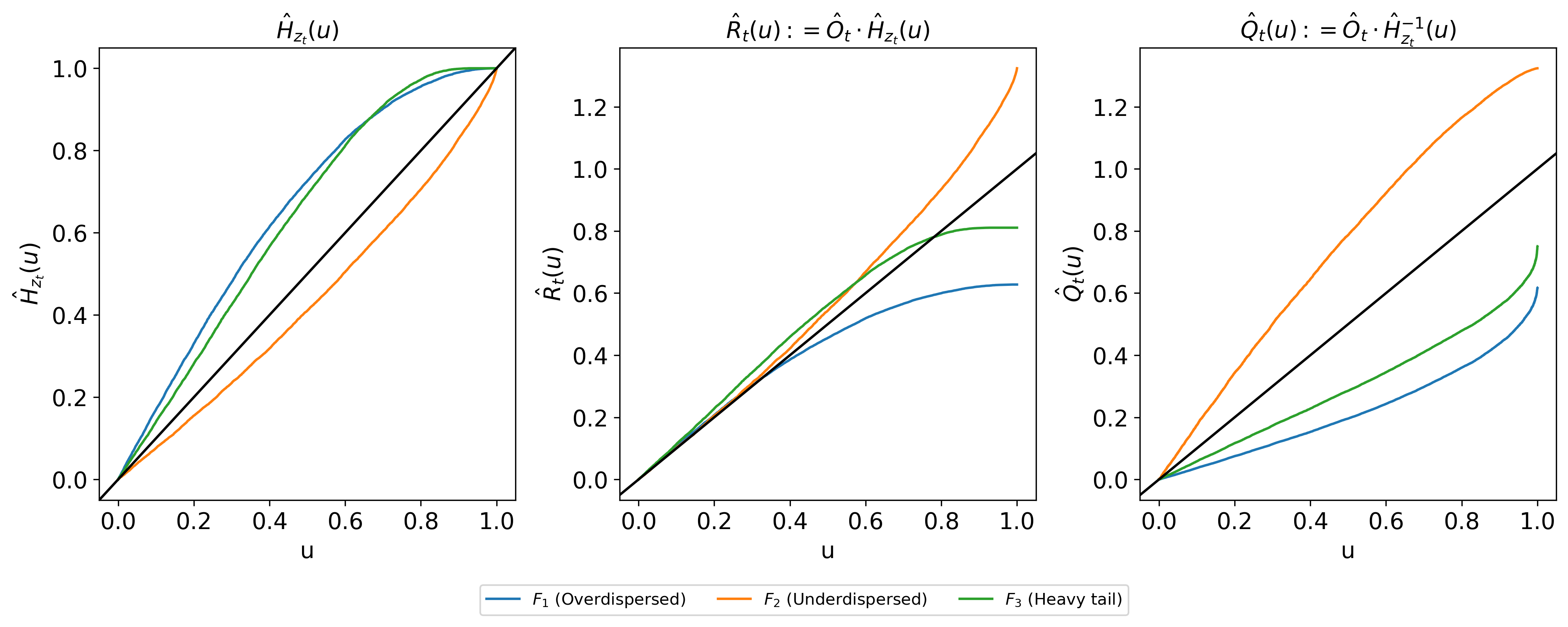}
    \caption{Empirical distribution function $\hat{H}_{z_t}$ of conditional PIT values (left), $\hat{R}_{t}(u)$ against $u$ (middle), and $\hat{Q}_{t}(u)$ against $u$ (right), for the three miscalibrated forecasts $F_1, F_2$, and $F_3$, with $t = 1.0$. The estimates of $O_t$ corresponding to the three forecasts are $\hat{O}_{t} = 0.63, \hat{O}_{t} = 1.34, \hat{O}_{t} = 0.82$ for $F_1, F_2, F_3$, respectively. }
    \label{fig:R_hat_Q_hat}
\end{figure}

To illustrate this, consider the following simple simulation experiment. We simulate $100,000$ standard normal observations $Y \sim N(0,1)$ and consider three forecasts:
\begin{itemize}
    \item Forecast 1 -- Overdispersed: $F_1 \sim N(0, 1.5)$
    \item Forecast 2 -- Underdispersed: $F_2 \sim N(0, 0.85)$
    \item Forecast 3 -- Heavy tailed: $F_3 \sim \mathrm{Student-}t(3)$
\end{itemize}
We investigate tail calibration at the threshold $t = 1.0$. Figure \ref{fig:R_hat_Q_hat} displays the distribution function of the conditional PIT values, as well as $\hat{R}_{t}(u)$ and $\hat{Q}_{t}(u)$ as a function of $u$. One can see that for the forecasts $F_1$ and $F_3$, the distribution functions of CPIT values (left panel) lie above the diagonal line, indicating the excess distributions are too heavy tailed, while the corresponding distribution function for $F_2$ indicates a tail that is too light. However, this information is not available from the plot of $\hat{R}_{t}(u)$ (middle panel), where the curves for $F_1$ and $F_2$ lie on the opposite side of the diagonal line to before, and the curve for the heavy tailed forecast $F_3$ crosses the diagonal, appearing reasonably well tail calibrated despite both components ($\hat{H}_{z_t}$ and $\hat{O}_t$) indicating that the forecast is substantially miscalibrated in the tails. 
On the other hand, the diagnostic plot based on $\hat{Q}_{t}(u)$ (right panel) avoids this issue, with the errors in $\hat{H}_{z_t}$ amplified by the errors in $\hat{O}_t$. 

\section{Model details}\label{sec:app_model}

We provide some more details on the models used in Section \ref{sec:applications}. Code to implement all of the methods can be found at \url{https://github.com/jakobwes/Enforcing-tail-calibration}.

\subsection{Ensemble Model Output Statistics}\label{sec:app_EMOS}

Ensemble model output statistics \citep[EMOS;][]{jewson_new_2004, GneitingEtAl2005} is a popular post-processing technique to recalibrate ensemble forecasts issued by a numerical weather prediction model. EMOS models the target variable, here wind speed $W$, in terms of a parametric distribution whose location and scale parameters depend on the mean and standard deviation of the ensemble forecast at the corresponding time and location. Following \cite{thorarinsdottir_probabilistic_2010}, we use a truncated normal distribution $N_0(\mu, \sigma)$ for wind speed. The location parameter $\mu$ is modelled as a linear function of the ensemble mean $m$, while the (log) scale parameter $\log \sigma$ is modelled linearly in terms of the ensemble standard deviation $s$; for simplicity, we suppress the dependence of $m$ and $s$ on the time and location in the notation. We also include the day of year ($doy$) as covariate to account for seasonal variations in forecast skill and biases:
\begin{align*}
    &W\ |\ m, s, doy \sim N_0(\mu, \sigma)\\
    &\mu = \alpha + \beta m + \lambda_{\mu, s} \sin \left( \tfrac{2 \pi\  doy}{365.25} \right) + \lambda_{\mu, c} \cos \left( \tfrac{2 \pi\  doy}{365.25} \right)\\
    &\log \sigma = \eta + \delta s + \lambda_{\sigma, s} \sin \left( \tfrac{2 \pi\  doy}{365.25} \right) + \lambda_{\sigma, c} \cos \left( \tfrac{2 \pi\  doy}{365.25} \right)
\end{align*}

The parameters to estimate are then $\theta = (\alpha, \beta, \eta, \delta, \lambda_{\mu,s}, \lambda_{\mu,c},\lambda_{\sigma,s}, \lambda_{\sigma,c})$. While the parameters could be estimated at each weather station individually, we use a semi-local post-processing approach \citep[e.g.][]{lerch_similarity-based_2016} to increase the amount of data available for model training, which is particularly relevant when forecasting extreme events, as discussed by \citet{WesselEtAl2024}. In this approach, individual EMOS models are fit to clusters of similar stations. This corresponds to location-wise fitting if one cluster per station is used, while smaller numbers of clusters allow data to be shared efficiently across stations. 

We follow \citet{WesselEtAl2024} and cluster based on the observed climatology. For this, we define $K$ features as the $1/(K+1), \dots, K/(K+1)$th quantiles of the observational distribution. Subsequently, $k$-means clustering is used to find clusters of similar stations. We use the ``elbow'' technique to select a cluster number and identify four distinct clusters corresponding broadly to coastal, more exposed coastal, inland and more exposed inland stations (see Figure \ref{fig:locations}). EMOS models are subsequently fit to the data of all stations in each of the clusters.

For parameter optimization we use closed-form CRPS and twCRPS expressions for a truncated normal distribution \citep{jordan_evaluating_2019, wessel_lead-time-continuous_2024}. We follow the practice of the \texttt{crch} R-package \citep{MessnerEtAl2016}. We use a Broyden-Fletcher-Goldfarb-Shanno (BFGS) quasi-Newton optimizer \citep{nocedal_numerical_2006} with empirical gradient estimates, falling back on a Nelder-Mead optimizer \citep{nelder_simplex_1965} should it fail. As optimization starting values, the results of a linear regression are used for the location parameters, whilst the scale intercept is initialized to the log-transformed standard error and all other parameters are set to zero. As the optimizers rely on empirical gradients, the (tail) miscalibration penalties can easily be incorporated into the objective functions.

\subsection{Distributional Regression Network}\label{sec:app_DRN}

Distributional regression networks \citep{rasp_neural_2018} are a nonlinear extension of EMOS. Wind speed is again assumed to follow a truncated normal distribution $N_0(\mu, \sigma)$, whose parameters are modelled not as linear functions, but in terms of feedforward neural networks. DRNs thus allow nonlinear dependencies between the covariate and the outcome to be captured, whilst maintaining the robustness that a parametric assumption offers. 

The neural networks have two hidden layers with 16 units each and rectified linear unit (ReLU) activation functions. As the scale parameter $\sigma$ is constrained to be positive, it is exponentially transformed after the final layer. We again use the ensemble mean, ensemble standard deviation as well as the sine and cosine transformed normalized day of year ($\tfrac{2\pi\ doy}{365.25}$) as covariates. One single DRN is fitted to all stations, however in order to make the post-processing spatially aware, the station index is used in a learned embedding layer, embedding location information in two dimensions, which are used as covariates. 

We use the closed-form expression of the CRPS for a truncated normal distribution for CRPS-based training. We use the Adam optimizer \citep{kingma_adam_2017}, a stochastic gradient descent method, for parameter optimization, and use minibatching with a large batch size of 2048. We train for a fixed number of 200 epochs and find good convergence in all cases. 

After CRPS-based training, we finetune the CRPS-pre-trained models using the (tail) calibration and twCRPS penalties for a threshold of $t=12.5$m/s, and values of $\gamma = 1.0, 2.5, 5.0, 10.0, 20.0$. We do not use minibatching to compute the CRPS and calibration scores during finetuning, since the calibration penalties rely on the entire set of forecasts and observations across the training dataset to calculate the empirical distribution of the PIT values or the $\hat{Q}$ values. While this can be estimated within batches, these estimates can be unstable for small batch sizes and high thresholds, where only a small amount of data is available. Instead, we pass through the whole training dataset to compute the CRPS and calibration losses, similar to the EMOS models for 50 steps. Whilst full training with the full batch and the calibration-penalized losses might also be considered, we decide to finetune CRPS-trained models as this speeds up convergence. We use again an Adam optimizer for this finetuning step.


\subsection{Conditional Generative Model}\label{sec:app_CGM}

Conditional generative models \citep[CGMs,][]{ chen_generative_2024} provide a non-parametric alternative to DRN and EMOS models. They construct a predictive distribution by transforming samples from a latent distribution, here standard Normal, using neural networks. The predictive distribution is then available via samples. CGMs do not rely on parametric assumptions directly and are able to represent a wide range of predictive distributions.

We fit one CGM for all stations, using again an embedding layer to encode spatial information in two dimensions. The CGM model architecture follows the structure proposed by \cite{chen_generative_2024}. As covariates, we again use the ensemble mean and standard deviation, sine and cosine transformed normalized day of year, as well as location embeddings. The CGM model broadly consists of two parts: one post-processing the ensemble mean and one accounting for the uncertainty by post-processing the latent noise and ensemble standard deviation. These are added together and transformed with a softplus activation to ensure positivity of the predicted wind speed values. We use two hidden layers with 16 units each adjusting the ensemble mean, together with the external covariates. Similarly, the latent Gaussian noise is scaled up by the ensemble standard deviation and transformed, together with the exogenous covariates, through two layers with 16 units again.

Similar to the DRN case, we pre-train using the CRPS and finetune with the twCRPS and (tail) calibration penalties. Since the CGM outputs a predictive distribution via samples, we rely on the sample expression of the CRPS for a discrete predictive distribution in \eqref{eq:crps_sample} for CRPS-based training. We use the Adam optimizer for parameter optimization and, similarly to the DRN case, use minibatching with a batch size of 2048. We train the models for a fixed number of 50 epochs and find good convergence. In each step, we construct a predictive distribution using an ensemble of 250 samples that we use to compute the sample CRPS.

After CRPS-based training, we finetune using the (tail) calibration penalties and twCRPS for a threshold of $t = 12.5$m/s and $\gamma = 1.0, 2.5, 5.0, 10.0, 20.0$. As for the CRPS, PIT values and the $\hat{Q}$ values for the miscalibration penalties are computed using the empirical distribution functions defined by the sample from the CGM. With a predictive distribution given through samples $\hat{y}_1, \dots, \hat{y}_n$ and an observation $y$, an estimate of the PIT value can be obtained through the normalized rank of $y$ among $\hat{y}_1, \dots, \hat{y}_n$. However, ranking as a mathematical operation is not differentiable, which is a challenge for stochastic gradient descent methods used for training neural networks, which rely on gradients of the loss function with respect to the neural network parameters. Thus, we approximate the ranking via smooth approximations to step functions. Specifically, we approximate the PIT using
\begin{equation}
\frac{1}{n}\left[\text{rank}(y, \{\hat{y}_1, \dots, \hat{y}_n\}) - 1\right] = \frac{1}{n}\sum_{i=1}^n \one\{\hat{y}_i \leq y\} \approx \frac{1}{n} \sum_{i = 1}^n \sigma\left(\frac{y - \hat{y}_i}{\nu}\right),
\end{equation}
where $\sigma(x) = \tfrac{1}{1+e^{-x}}$ is the sigmoid function and $\nu$ is a hyperparameter that ensures the quality of the approximation. For smaller values of $\nu$ the approximation becomes more exact, whilst the objective remains smoother for larger values of $\nu$. We choose $\nu = 1/100$.

As for the DRN model, batch-wise training is not possible when penalizing miscalibration, since the miscalibration terms depend on the forecast and observations in the whole training dataset. However, fitting the CGM on the whole training dataset is computationally prohibitive, since it becomes expensive to calculate the sample-based CRPS; this scales quadratically with the ensemble size $M$ and is of complexity $\mathcal{O}(N M + N M^2)$ for $N$ training samples, thus becoming infeasible for large batch sizes and ensemble sizes. To address this, we employ the fair CRPS for training (see \eqref{eq:fcrps_sample} in Appendix \ref{app:samplescores}), as proposed in \citet{lang_aifs-crps_2024}. The fair CRPS is essentially an unbiased estimator for the population CRPS, and is therefore less sensitive to the sample size; it remains computationally tractable for an ensemble size of $M = 50$. We thus finetune with losses of the form $fCRPS + \gamma TMCB$, using the whole training dataset for each gradient estimation. We also use a fair twCRPS. \citet{lang_aifs-crps_2024} find that the fair CRPS is degenerate when all but one ensemble members have the same value, which can make training of the CGM instable. However, this did not cause any problems here, due to the higher floating point precision used. Alternatively, one could use the almost-fair CRPS as introduced in \citet{lang_aifs-crps_2024}. 
The (tail) calibration penalties are computed as before using $M = 250$ samples.  

\section{Scoring rules for forecast samples}\label{app:samplescores}

When $F$ has finite first moment, the CRPS defined in \eqref{eq:crps} can alternatively be written as 
\begin{equation}
    \mathrm{CRPS}(F, y) = \E | X - y | - \frac{1}{2} \E |X - X^{\prime} |
\end{equation}
where $X, X^{\prime} \sim F$ are independent \citep{GneitingRaftery2007}. The twCRPS can similarly be expressed as 
\begin{equation}
    \mathrm{twCRPS}(F, y) = \E | v(X) - v(y) | - \frac{1}{2} \E |v(X) - v(X^{\prime}) |,
\end{equation}
where $v$ is an anti-derivative of the weight function $w$, that is, $v(x) - v(x^{\prime}) = \int_{x^{\prime}}^{x} w(z) \dd z$ for all $x, x^{\prime} \in \R$ \citep{taillardat_evaluating_2023,AllenEtAl2023b}. When $w(x) = \one\{x > t\}$, for some threshold $t \in \R$, one choice for $v$ is $v(x) = \max\{x, t\}$.

When the forecast $F$ is a discrete predictive distribution, the CRPS and twCRPS can be obtained by replacing the expectations in the above expressions with sample means. For example, suppose $F$ is the empirical distribution function corresponding to a sample of points $\hat{y}_{1}, \dots, \hat{y}_{M} \in \R$, that is, $F(x) = \sum_{i=1}^{M} \one\{\hat{y}_i \leq x\} / M$. In this case, the CRPS becomes
\begin{equation}\label{eq:crps_sample}
    \mathrm{CRPS}(F, y) = \frac{1}{M} \sum_{i=1}^{M} |\hat{y}_{i} - y| - \frac{1}{2M^{2}} \sum_{i=1}^{M} \sum_{j=1}^{M} |\hat{y}_{i} - \hat{y}_{j}|,
\end{equation}
and the twCRPS becomes \citep{AllenEtAl2023}
\begin{equation}
        \mathrm{twCRPS}(F, y) = \frac{1}{M} \sum_{i=1}^{M} |v(\hat{y}_{i}) - v(y)| - \frac{1}{2M^{2}} \sum_{i=1}^{M} \sum_{j=1}^{M} |v(\hat{y}_{i}) - v(\hat{y}_{j})|.
\end{equation}
These sample versions of the CRPS and twCRPS are used to evaluate the conditional generative model forecasts in Section \ref{sec:applications}.

If the sample is to be treated as random draws from an underlying distribution, rather than a predictive distribution in its own right, then the sample CRPS can be adjusted to account for biases that arise due to the sample size. \cite{ferro_effect_2008} propose the fair CRPS (fCRPS) as 
\begin{equation}\label{eq:fcrps_sample}
    \mathrm{fCRPS}(F, y) = \frac{1}{M} \sum_{i=1}^{M} |\hat{y}_{i} - y| - \frac{1}{M(M-1)} \sum_{i=1}^{M} \sum_{j=1}^{M} |\hat{y}_{i} - \hat{y}_{j}|,
\end{equation}
which essentially corresponds to an unbiased estimator of the population CRPS for the distribution underlying the samples \citep{ferro_fair_2014}. A fair threshold-weighted CRPS can be obtained analogously. We use the fair CRPS for finetuning the conditional generative models in Section  \ref{sec:applications}, where larger training datasets require us to reduce the amount of generated samples $M$ from the model. 

\section{Regularizing with conditional PIT values}
\label{app:sec:penalizing-cpit}
Tail miscalibration has traditionally also be assessed using histograms of conditional PIT (CPIT) values \citep{AllenEtAl2023b, MitchellWeale2023}. These are values 
\begin{equation}
    z_{i,t} = \frac{F_{i}(y_{i}) - F_{i}(t)}{1 - F_{i}(t)}, \quad \text{for} \quad i \in \Ii_{t},
\end{equation}
where $\Ii_{t} = \{ i \in \{ 1, \dots, n \} : y_{i} > t \}$ is the set of indices for which the outcome exceeds the threshold. If the exceedance distribution is probabilistically calibrated the empirical distribution function of conditional PIT values $\hat{H}_{z_t}$ should be (close to) uniform. Unlike the combined ratio in the $\hat{Q}$ values, however, the CPIT values do not account for (mis-)calibration of the occurrence of extreme events, but only for the probabilistic calibration of the intensity distribution (see eq. \eqref{eq:decomposition}). 

Similarly to the MCB and TMCB one can also obtain a measure of miscalibration in the CPIT values, by penalizing the distance of the empirical distribution of conditional PIT values $\hat{H}_{z_t}$ to uniformity:
\begin{equation}\label{eq:CPIT_MCB}
\text{CPIT-MCB} = \int_0^1 |\hat{H}_{z_t}(u) - u |\ \mathrm{d} u 
\end{equation}
This can be penalized during training similar to the MCB and TMCB. 

\begin{figure}[h]
    \centering
    \includegraphics[width=\linewidth]{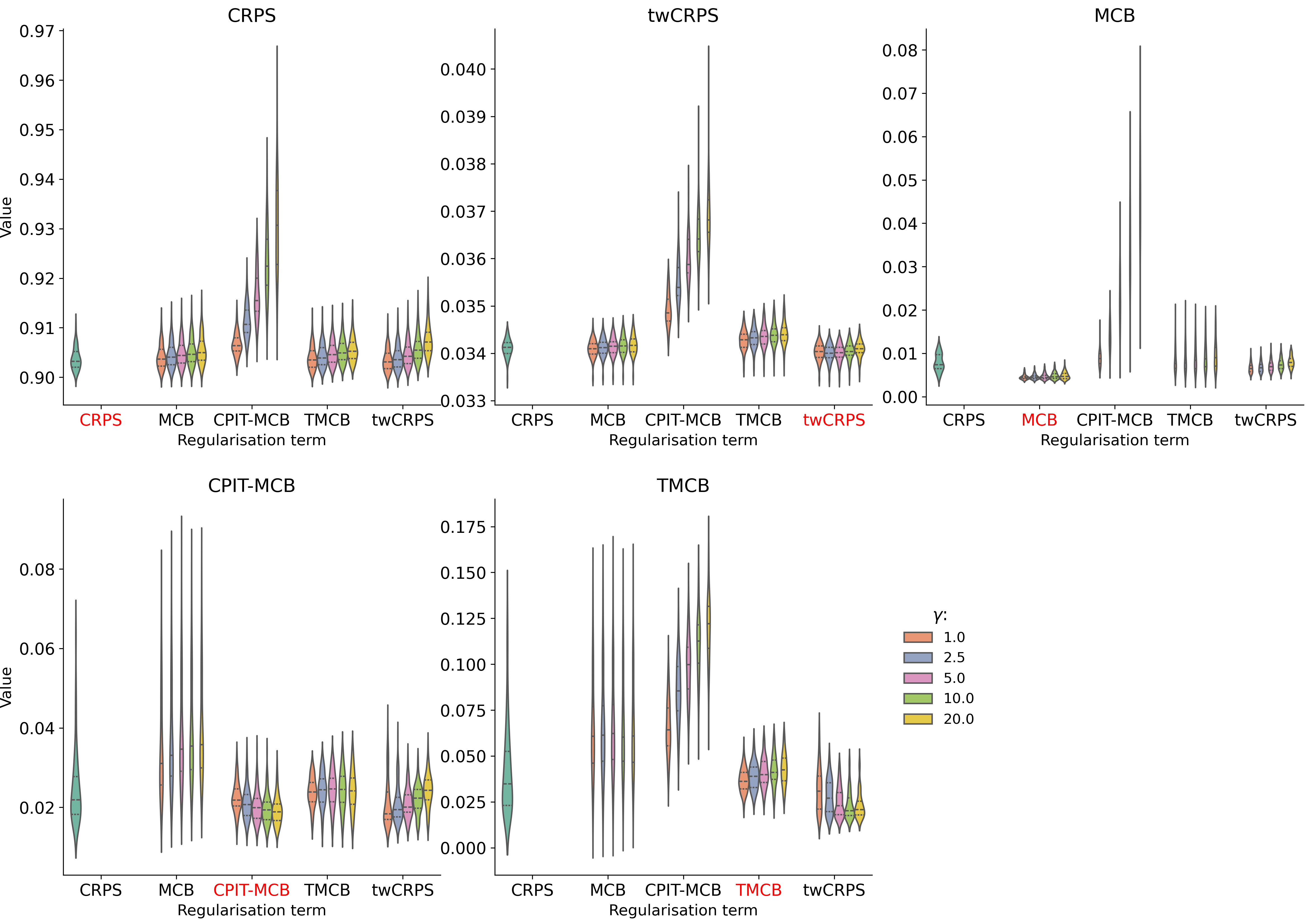}
    \caption{Violin plot of metrics for the 100 DRN models trained using different regularization terms. Different panels correspond to different evaluation metrics, while the x-axes labels correspond to the different regularization terms. Red x-axis labels correspond to when the evaluation metric is the same as the metric used for regularization.}
    \label{fig:DRN_CPIT}
\end{figure}

\begin{figure}[h]
    \centering
    \includegraphics[width=\linewidth]{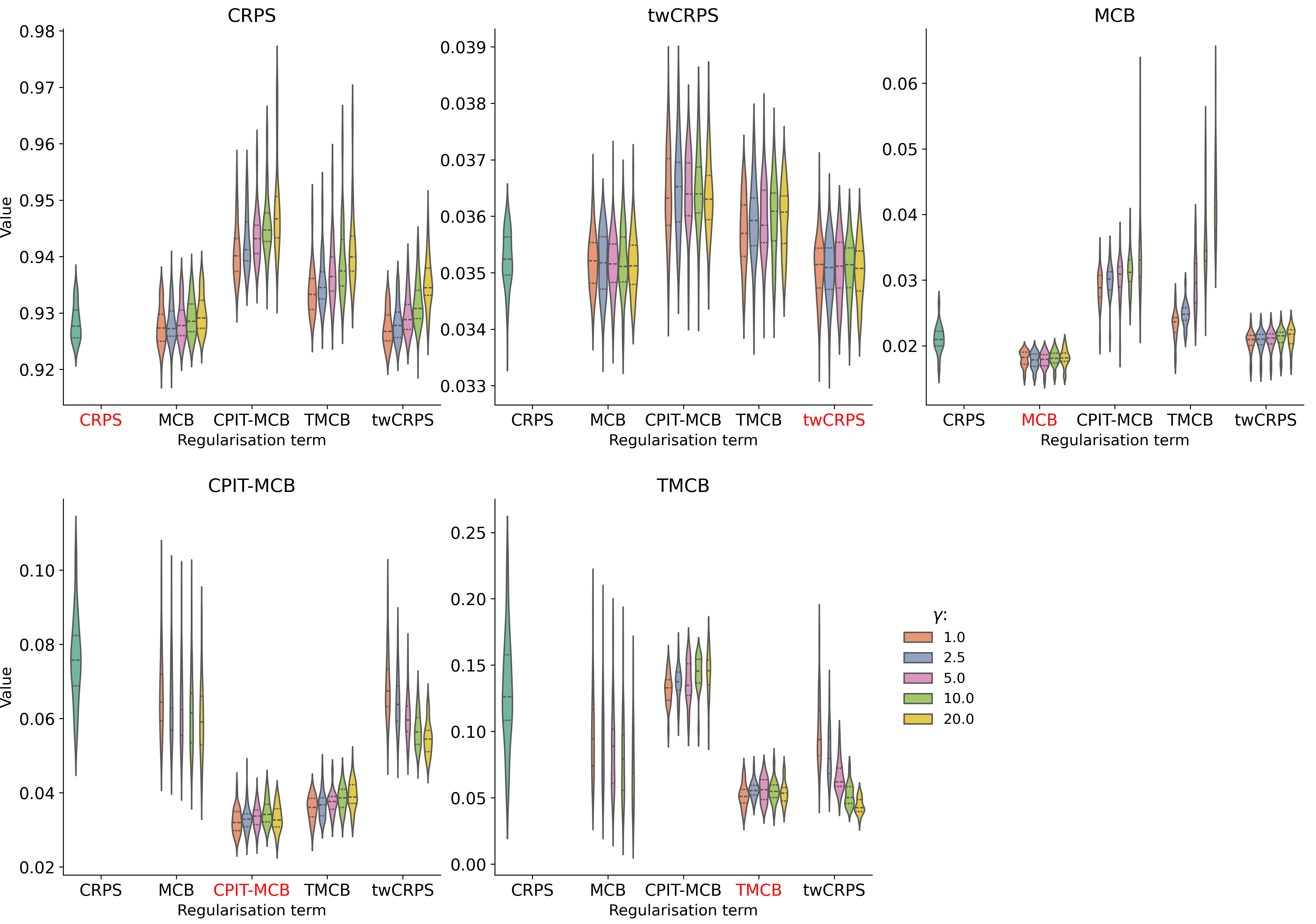}
    \caption{Violin plot of metrics for the 100 CGM models trained using different regularization terms. Different panels correspond to different evaluation metrics, while the x-axes labels correspond to the different regularization terms. Red x-axis labels correspond to when the evaluation metric is the same as the metric used for regularization.}
    \label{fig:CGM_CPIT}
\end{figure}

Figure ~\ref{fig:DRN_CPIT} and Figure ~\ref{fig:CGM_CPIT} contain the scores for the DRN and CGM models for training with all the tail calibration penalties including the CPIT-MCB. This includes the results in Figures \ref{fig:DRN_2} and \ref{fig:CGM_2}, as well as additional results for CPIT-MCB regularization and evaluation. Results for the EMOS have been omitted for brevity, but are largely consistent. CPIT-MCB penalization improves the CPIT-MCB for both model types. This, however, comes at a large cost in terms of CRPS, MCB and also TMCB where the scores deteriorate strongly. The CPIT-MCB penalization improves the calibration of the exceedance distribution. However, it does so by shifting large amounts of probability mass beyond the threshold and tends to strongly overpredict threshold exceedances. The TMCB and CRPS both improve the CPIT-MCB, but also lead to improvements in terms of TMCB.

This behavior indicates the need to assess both the calibration of the threshold exceedance (occurrence) forecast as well as the calibration of the exceedance distribution (intensity) when evaluating calibration of extreme event forecasts. This can be understood as a calibration version of the Forecaster's dilemma discussed by \citet{LerchEtAl2017}, who argue for the need to evaluate both intensity and occurrence of extreme events forecasts jointly. The TMCB addresses this by adjusting the CPIT values with the occurrence ratio.

\end{document}